\documentclass[11pt,a4paper]{article}
\usepackage[...]{youngtab}
\usepackage{appendix}
\def\Bbb{\mathbb}
\usepackage{amsfonts}
\usepackage[usenames,dvipsnames]{xcolor}
\usepackage{color}
\definecolor{Red}{rgb}{1,0,0}
\def \BZ{\mbox{$\Bbb Z$}}

\def \BC{\mbox{$\Bbb C$}} 
\def \BP{\mbox{$\Bbb P$}}

\def \P3{\mathbb{P}^3}
\def \Ba{{\cal B}_1}
\def \Bb{{\cal B}_2}
\def \Bc{{\cal B}_3}
\def \Bd{{\cal B}_4}
\def \Ca{{\cal C}_1}
\def \Cb{{\cal C}_2}
\def \Cc{{\cal C}_3}
\def \Cd{{\cal C}_4}
\def \Ce{{\cal C}_5}

\def \Db{{\cal D}_2}
\def \Ea{{\cal E}_1}
\def \Eb{{\cal E}_2}
\def \Ec{{\cal E}_3}
\def \Ed{{\cal E}_4}
\def \Fa{{\cal F}_1}
\def \Fb{{\cal F}_2}
\def \susyN{\mathcal{N}}
\def \toricG{{\cal{G}}}

\newcommand{\sid}{\begin{equation}}
\newcommand{\sidd}{\end{equation}}
\usepackage{pgf}
\usepackage{tikz}
\usepackage[utf8]{inputenc}
\usetikzlibrary{arrows,automata}
\usetikzlibrary{positioning}
\tikzset{state/.style={rectangle, rounded corners, draw=black, very thick, minimum height=2em, inner sep=2pt, text centered,},}
\begin{document}
\begin{center}
\Large{\textbf{Embedding and partial resolution of complex cones over Fano threefolds}}
\end{center}
\vspace{0.2in}
\textbf{Siddharth Dwivedi}
\vspace{0.1in}
\newline
Department of Physics, Indian Institute of Technology Kanpur,\\
Kanpur 208016, India
\vspace{0.1in}
\newline
E-mail: sdwivedi@iitk.ac.in
\vspace{0.3in}
\newline
\textbf{Abstract}:
This work deals with the study of embeddings of toric Calabi-Yau fourfolds which are complex cones over the smooth Fano threefolds. In particular, we focus on finding various embeddings of Fano threefolds inside other Fano threefolds and study the partial resolution of the latter in hope to find new toric dualities.

We find many diagrams possible for many of these Fano threefolds, but unfortunately, none of them are consistent quiver theories. We also obtain a quiver Chern-Simons theory which matches a theory known to the literature, thus providing an alternate method of obtaining it. 
\newpage

\tableofcontents

\section{Introduction}
\label{sec1}
The relation between the gauge theory and geometry has been an active area of research for a long time. $D3$ and $M2$ branes have been extensively used for probing various space-time singularities. The gauge theory living on the worldvolume of the branes is influenced by the geometry of the space-time singularity probed by these branes. Thus  the information about the geometry, as a result of probing, is transformed into the gauge theory data.

The study of $D3$-branes probing the trivial flat space $\BC^3$ has led to one of the most important developments in string theory, called the AdS/CFT correspondence \cite{Maldacena:1997re}. The system of $N$ coincident $D3$-branes in flat space can be viewed as both type-IIB string theory on $AdS_5 \times S^5$ and also as (3+1)-dimensional $\susyN=4$ supersymmetric $SU(N)$ Yang-Mills theory \cite{Maldacena:1997re}. It was found that by taking the orbifolds $\BC^3 / \Gamma$, where $\Gamma \subset SU(3)$ is a subgroup of $SU(3)$, the gauge theories with less number of supersymmetries can be constructed \cite{Kachru:1998ys, Lawrence:1998ja, Hanany:1998sd}. An important class of such orbifolds is the abelian orbifold for which $\Gamma$ is of the form $\left( \BZ_m \times \BZ_n \right)$. These abelian orbifolds of $\BC^3$ reduce the supersymmetry to $\susyN=1$ \cite{Douglas:1997de}. The matter content of the gauge theories which arise on the $D3$-branes as a result of probing orbifolds of $\BC^3$, can be represented in terms of a quiver diagram. A quiver diagram consists of some nodes and oriented arrows connecting the nodes \cite{Douglas:1996sw}. The nodes represent the gauge groups of the gauge theory and the oriented arrows between two nodes represent the bifundamental chiral multiplets. An arrow starting and ending on the same node represents an adjoint field. Moreover, the superpotential of these gauge theories is a subset of closed loops (formed by the oriented arrows of the quiver) in the quiver diagram. Such a gauge theory is called as quiver gauge theory.

The abelian orbifolds of $\BC^3$ are examples of a special class of manifolds, called toric Calabi-Yau threefolds ($CY_3$) \cite{fulton1993introduction, Aharony:1997bh, Leung:1997tw}. A general toric $CY_3$ can also be constructed by taking a real cone over a 5-dimensional Sasaki-Einstein manifold \cite{Martelli:2004wu, Martelli:2005tp,boyer2008sasakian}. The geometrical information of a toric Calabi-Yau manifold is contained in its toric diagram. The toric diagram for a toric $CY_3$ is a 2-dimensional diagram which is a convex lattice polygon drawn on $\BZ^2$ lattice. The vertices of the toric diagram can be encoded in the columns of a matrix, called as toric data, which we shall denote as $\toricG$. These toric $CY_3$ have a cone type singularity. A stack of $D3$-branes  placed transversely at the tip (or the singularity) of a toric $CY_3$ gives rise to $\susyN=1$ quiver gauge theory on the worldvolume of $D3$-branes. The toric nature of the underlying Calabi-Yau geometry restricts the gauge group of the quiver gauge theory of the form $\prod U(N)^G$, where $G$ is the number of gauge group factors or the number of nodes in the quiver diagram. Further, the superpotential $W$ of these gauge theories is toric, which means that every field should appear exactly twice in $W$, once in a positive term and once in a negative term.

It is an interesting exercise to determine the gauge group and matter content $\{X_i \}$'s (or the quiver diagram) and the superpotential of the $\susyN=1$  quiver gauge theory for a given toric $CY_3$. The reverse problem is to determine the geometry of the toric $CY_3$ corresponding to the quiver gauge theory residing on $D3$-branes transverse to the Calabi-Yau. In the literature, the method of obtaining toric data from quiver gauge theory is called the \emph{forward algorithm} and the reverse case, i.e. obtaining quiver gauge theory data from the toric data is called the \emph{inverse algorithm} \cite{Feng:2000mi, Feng:2001xr}.

The forward algorithm starts with the quiver gauge theory and computes the toric data of the underlying Calabi-Yau singularity. In forward algorithm, the interactions $\left(W\right)$ and the matter content (quiver diagram) of a quiver gauge theory give the $F$-term and $D$-term equations \cite{Feng:2000mi}. These are used to compute the so-called matrices $Q_F$ and $Q_D$ \cite{Feng:2000mi}. These matrices can be concatenated together in a larger matrix $Q$, called as total charge matrix \cite{Feng:2000mi} and is given as:
\begin{equation}
Q =  \left(
\begin{array}{c}
 Q_F \\ \hline
 Q_D
\end{array}
\right)~.
\end{equation}
The nullspace or the cokernel of the total charge matrix $Q$ gives the toric data $\left(Q.\toricG^t=0\right)$. For a toric $CY_3$, the columns of toric data turn out to be three-dimensional vectors but the Calabi-Yau condition requires these vectors to be coplanar. Thus the tip of these vectors can be drawn as a convex polygon in $\BZ^2$ lattice, which is the required toric diagram of the toric $CY_3$. In other words, this toric diagram can be drawn in a two-dimensional plane whose vertices are always integer points.

Reversing the steps of the forward algorithm, one can in principle obtain quiver gauge theories from the Calabi-Yau threefold toric data. This procedure is called inverse algorithm \cite{Feng:2000mi, Feng:2001xr}. However the inverse algorithm has ambiguities \cite{Feng:2000mi}. Hence to perform the inverse algorithm on a toric data $\toricG$, we have to determine the multiplicity of columns of $\toricG$ as well as the $Q_F$ matrix and the $Q_D$ matrix by some other method. 

For any general toric Calabi-Yau threefold, the charge matrices $Q_F$ and $Q_D$ and the correct multiplicity in toric data can be obtained using the method of partial resolution \cite{Douglas:1997de, Morrison:1998cs, Feng:2000mi, Beasley:1999uz} on $\left( \BZ_m \times \BZ_n \right)$ orbifold singularities of $\BC^3$. In this method, a given Calabi-Yau threefold is embedded into $\BC^3 /  \left( \BZ_m \times \BZ_n \right)$, where $m$ and $n$ are the smallest integers such that the toric diagram of the orbifold contains the toric diagram of the given $CY_3$. The dual quiver gauge theory for the $\BC^3 / \left( \BZ_m \times \BZ_n\right) $ is well known. The correct multiplicity of $\toricG$ and corresponding $Q_F$ and $Q_D$ charge matrices of $CY_3$ can be obtained using the steps of partial resolution method \cite{Feng:2000mi} by removing appropriate points from the toric diagram of abelian orbifold of $\BC^3$. It is important to mention that the method of partial resolution can be applied for any general toric $CY_3$ to obtain the possible quiver gauge theories. However, as the values of $m$ and $n$ become large, the computations become difficult.

Motivated by the $D3$-branes probing geometrical singularities, string theorists started to use $M2$-branes for probing the toric Calabi-Yau fourfold ($CY_4$) singularities. However a general structure of the underlying (2+1)-dimensional conformal field theory ($CFT_3$) was unknown for a long time. It was realized later that by introducing the Chern-Simons terms, one can construct (2+1)-dimensional field theories with more than $\susyN=3$ supersymmetry. This idea was used by Bagger and Lambert \cite{Bagger:2006sk, Bagger:2007jr, Bagger:2007vi} and by Gustavsson \cite{Gustavsson:2007vu, Gustavsson:2008dy} and various higher supersymmetric ($\susyN=3,4,5,6$) Chern-Simons theories were constructed \cite{Aharony:2008ug, Gaiotto:2008sd, Hosomichi:2008jd, Hosomichi:2008jb, Benna:2008zy, Schnabl:2008wj, Bergshoeff:2008bh, Bagger:2008se, Jafferis:2008qz}. It led to the understanding of many $AdS_4/CFT_3$ duals pairs.

In the pioneering work by Aharony, Bergman, Jafferis and Maldacena (ABJM) \cite{Aharony:2008ug}, a two-node quiver gauge theory was conjectured to be dual to coincident $M2$-branes probing $\BC^4/ \BZ_K$ orbifold. This gauge theory is a $\susyN=6$ supersymmetric $U(N) \times U(N)$ Chern-Simons theory with integer Chern-Simons levels $(k, -k)$. This theory is also famous by the name of ABJM theory \cite{Aharony:2008ug} which has the gravity dual as $M$-theory on $AdS_4 \times S^7 / \BZ_k$. For $k=1$, this theory describes $M2$-branes probing flat $\BC^4$.

More general quiver Chern-Simons theories with less number of supersymmetries ($\susyN=2$) were constructed by placing $M2$-branes at the tip of singular Calabi-Yau fourfolds. These $CY_4$ have the base manifold $Y$, which is a seven-dimensional Sasaki-Einstein manifold (analogous to the five-dimensional base in the $CY_3$ case). These theories were conjectured to be dual to $M$-theory on $AdS_4 \times Y$. 

Note that the matter content of these $\susyN=2$ quiver Chern-Simons theories can be encoded in the usual quiver diagrams. However, to completely specify a quiver Chern-Simons theory, we also have to indicate the integer Chern-Simons levels corresponding to each gauge group factor. This is achieved by putting the integers on the corresponding nodes of the quiver diagram. We also require these quiver Chern-Simons theories to satisfy the two toric conditions mentioned earlier: gauge group is $\prod U(N)^G$, where $G$ is the number of nodes in the quiver and $W$ is toric (every field should appear exactly twice with alternate signs). There is an additional constraint on the Chern-Simons levels of this theory from vacuum equations for $\susyN=2$ superconformal Chern-Simons with gauge group $U(1)^G$:
\begin{equation}
\sum_{i=1}^G k_i = 0 ~.
\end{equation}
The sum of Chern-Simons levels equals zero will be necessary to get the Calabi-Yau to be fourfold \cite{Davey:2009sr}.

Similar to the forward algorithm for $\susyN=1$ quiver gauge theories, an extended version incorporating the additional input of Chern-Simons levels was given in \cite{Ueda:2008hx,Hanany:2008gx}. This algorithm can be used to obtain the toric data ($\toricG$) of Calabi-Yau fourfolds starting from $\susyN=2$ quiver Chern-Simons theories.

However it should be mentioned here that unlike the abelian orbifolds of $\BC^3$, the (2+1)-dimensional quiver gauge theories (and hence $Q_F$, $Q_D$) for general orbifolds $\left( \BZ_{n_1} \times \BZ_{n_2} \times \BZ_{n_3} \right)$ of $\BC^4$ are not known. In \cite{Benishti:2009ky}, the quiver theories for $\BC^4 / \left( \BZ_2 \right)^3$ were obtained, but a general method for constructing quiver Chern-Simons theories for an arbitrary orbifold of $\BC^4$ is not clear. Hence the method of partial resolution of $\BC^4$ orbifolds to obtain the quiver theories for any given Calabi-Yau fourfold is not applicable. 

An interesting class of toric Calabi-Yau manifolds are those which are constructed by taking complex cones over Fano varieties as base. Fano varieties in $d$-complex dimensions are usually called Fano $d$-folds. The important feature of these Fanos is that if we take a complex cone over a Fano $d$-fold, the resulting manifold will be a Calabi-Yau $(d+1)$-fold. In particular, if the Fano variety is toric, the corresponding Calabi-Yau will also be toric.

There are five smooth toric Fano twofolds in two complex dimensions, which are commonly known as zeroth Hirzebruch surface $\mathbb{F}_0$, and the del-Pezzo surfaces $dP_0$, $dP_1$, $dP_2$, $dP_3$ in the literature \cite{Feng:2000mi}. Taking the complex cone over a smooth toric Fano twofold will give a toric $CY_3$. The quiver gauge theories corresponding to the complex cones over these five smooth Fano twofolds was obtained in \cite{Feng:2000mi} by the partial resolutions of the $\left(\BZ_3 \times \BZ_3\right)$ orbifold of $\BC^3$. 

In three complex dimensions, there are 18 smooth toric Fano threefolds \cite{watanabe1982classification, batyrev1982toroidal} (with nomenclature as used in \cite{Davey:2011mz}): $\mathbb{P}^3$, ${\cal{B}}_1$, ${\cal{B}}_2$, ${\cal{B}}_3$, ${\cal{B}}_4$, ${\cal{C}}_1$, ${\cal{C}}_2$, ${\cal{C}}_3$, ${\cal{C}}_4$, ${\cal{C}}_5$, ${\cal{D}}_1$, ${\cal{D}}_2$, ${\cal{E}}_1$, ${\cal{E}}_2$, ${\cal{E}}_3$, ${\cal{E}}_4$, ${\cal{F}}_1$, ${\cal{F}}_2$.
Taking a complex cone over each of these will give a toric $CY_4$. The quiver Chern-Simons theories corresponding to 14 of these smooth toric Fano threefolds were obtained in \cite{Davey:2011mz} using the forward algorithm approach. The quiver Chern-Simons theories for the remaining four toric Fanos, i.e., Fanos $\BP^3$, $\Ba$, $\Bb$ and $\Bc$ was obtained in \cite{Dwivedi:2011zm} by analyzing the patterns of $Q_F$, $Q_D$ charge matrices and applying inverse algorithm. There is another quiver Chern-Simons theory known for Fano $\Ba$ which was obtained in \cite{Phukon:2011hp} and is different from the theory obtained in \cite{Dwivedi:2011zm}.

As already mentioned, the $Q_F$ and $Q_D$ for quiver Chern-Simons theories corresponding to general orbifolds of $\BC^4$ are not known. As a result, obtaining the quiver Chern-Simons theory for a given toric $CY_4$ by embedding its toric diagram into a bigger toric diagram of an abelian orbifold of $\BC^4$ and applying partial resolution, is not possible. However, we can still perform partial resolution on a toric $CY_4$, if we know the corresponding quiver Chern-Simons theory, i.e., if we know the $Q_F$ and $Q_D$ charge matrices. We shall explain the method of partial resolution in the next section.

It is an interesting exercise to obtain the embeddings of toric diagram of the complex cones over smooth toric Fano threefolds, into a larger toric diagram of some other Calabi-Yau fourfold where the corresponding quiver gauge theory for the latter is known. Some of these embeddings were discussed in \cite{Phukon:2011hp}. Using the methods of higgsing \cite{Agarwal:2008yb, Davey:2009qx} and unhiggsing \cite{Benishti:2009ky}, it was shown in \cite{Phukon:2011hp}, that the toric diagram of the complex cones over Fano threefolds can be embedded into Calabi-Yau fourfolds which may or may not be complex cones over Fano threefolds.  We also made a small progress in \cite{Dwivedi:2012qa}, where we studied the partial resolution of Fanos $\Ba$, $\Bb$, $\Bc$, $\Bd$. We succeeded in finding the toric diagram of Fano $\BP^3$ embedded inside the toric diagrams of Fano $\Bb$ and Fano $\Bc$. Moreover, performing partial resolution of the Fano $\Bb$ and Fano $\Bc$, whose quiver Chern-Simons theories are known in \cite{Dwivedi:2011zm}, we also obtained the quiver Chern-Simons theory for Fano $\BP^3$ \cite{Dwivedi:2012qa}. This quiver gauge theory for Fano $\BP^3$ matched with that obtained in \cite{Dwivedi:2011zm}, thus justifying the result of \cite{Dwivedi:2011zm}. 

In this work, we study the embedding of complex cones over the 18 smooth toric Fano threefolds, to find if they are embedded inside the complex cones of any other Fano threefold. Note that some of these embeddings such as embedding of Fano ${\cal{E}}_1$ inside Fano ${\cal{F}}_2$; Fano ${\cal{B}}_4$ inside Fano ${\cal{D}}_2$ and Fano $\Cd$; Fano ${\cal{C}}_3$ inside Fano ${\cal{E}}_3$ have been already done in \cite{Phukon:2011hp}. Moreover, once we have obtained a possible embedding, we apply partial resolution approach to obtain the $Q_F$ and $Q_D$ charge matrices. Then we apply inverse algorithm on these charge matrices to extract the possible quiver diagrams.  

The plan of the paper is as follows. In section \ref{sec2}, we briefly review partial resolution method and the steps which we are going to follow in the subsequent sections. In sections \ref{sec3}, \ref{sec4}, \ref{sec5}, \ref{sec6}, \ref{sec7}, \ref{sec8}, \ref{sec9}, we will discuss about the embeddings of Fano $\BP^3$, Fano $\Ba$, Fano $\Bd$, Fano $\Ca$, Fano $\Cc$, Fano $\Db$ and Fano $\Ea$ respectively into other Fano threefolds. We will also apply partial resolution to see if we can get quiver diagrams. We finally conclude in section \ref{sec10}.  
\section{Partial Resolution method}
\label{sec2}
We know that starting from a given toric Calabi-Yau fourfold, one can in principle, obtain the quiver Chern-Simons theory using the inverse algorithm. One major drawback of the inverse algorithm is that it suffers from subtle ambiguities \cite{Feng:2000mi} which we have listed below:
\begin{enumerate}
\item Two toric datas ${\cal G}$
 and ${\cal G}'$ are considered equivalent if they are related by any
$GL(4,{\BZ})$ transformation ${\cal T}$, that is, ${\cal G}={\cal T}.{\cal  G}'$. This 
leads to a huge pool of possible nullspace of ${\cal G}$ - namely, the charge matrix 
$Q$ satisfying $Q .{\cal G}^t=0$ can be many. Moreover, even with the knowledge of $Q$, there is no way to identify which rows of $Q$ form $Q_F$ and which rows form $Q_D$.
\item The multiplicity of the points in toric diagram gives the toric data
with repeated columns but they represent the same Calabi-Yau fourfolds.
A priory, it is not clear which points with what
multiplicity in the toric diagram to be taken. A different multiplicity will lead to a different charge matrix $Q$.
\end{enumerate}
The important step in inverse algorithm is to obtain the charge matrices $Q_F$ and $Q_D$ from the toric data. The quiver diagram, $W$ and Chern-Simons levels can then be obtained by reversing the steps of forward algorithm. Finding $Q_F$, $Q_D$ is the difficult part in the inverse algorithm. In the context of Calabi-Yau threefolds, there is a method to obtain the $Q_F$ and $Q_D$ charge matrices starting from any given toric $CY_3$. This method is known as partial resolution \cite{Feng:2000mi} and we will discuss it in this section.

Note that these ambiguities in inverse algorithm also give rise to an interesting duality called as toric duality \cite{Feng:2000mi}. Because of the ambiguities in choosing $Q_F$ and $Q_D$, we may get more than one quiver gauge theory corresponding to the same toric data of Calabi-Yau \cite{Feng:2000mi}. This suggests the presence of an underlying duality between such quiver gauge theories which is called the toric duality. The different quiver gauge theories which correspond to the same toric data are called toric duals or phases. Toric duality was first discovered in \cite{Feng:2000mi} while studying the partial resolution of $\BC^3 / \left(\BZ_3 \times \BZ_3\right)$ giving complex cones over del Pezzo surfaces, which are also known as Fano twofolds. The toric duality has been analyzed in many works \cite{Feng:2001xr, Feng:2002zw, Feng:2002fv}. Examples of various toric dual quiver gauge theories (or phases) have been discussed in \cite{Feng:2001xr, Davey:2009sr, Davey:2009qx}. 

\subsection{Partial resolution}
In the method of partial resolution, we start with a given toric Calabi-Yau threefold, say $CY_3^1$ with the toric data ${\cal{G}}_1$, whose charge matrices $Q_F^1$ and $Q_D^1$ needs to be found. For any toric Calabi-Yau threefold, the matrix ${\cal{G}}$ will have three rows, of which all the entries in the first row will be 1. This is because the Calabi-Yau condition requires that the tip of each three-dimensional vector (given by the column vectors in ${\cal{G}}$) lies in the same hyperplane. Thus the typical toric data in our case will be given as:
\begin{equation}
{\cal{G}}_1 = 
\left(
\begin{array}{ccccc}
 p_1 & p_2 & p_3 & \ldots & p_{\alpha } \\ \hline
 1 & 1 & 1 & \ldots & 1 \\
 n_{21} & n_{22} & n_{23} & \ldots & n_{2 \alpha } \\
 n_{31} & n_{32} & n_{33} & \ldots & n_{3 \alpha } 
\end{array}
\right)~,
\label{Toric-data-G1}
\end{equation}
where the $n_i$'s are all integers. Note that some of the columns in ${\cal{G}}_1$ may be repetitive which does not make any difference as far as the $CY_3^1$ is concerned. Consider another toric Calabi-Yau threefold, say $CY_3^2$ whose corresponding charge matrices (and hence quiver gauge theory) are already known. Further, we want this $CY_3^2$, with toric data ${\cal{G}}_2$, to be such that the toric diagram of $CY_3^1$ is embedded inside $CY_3^2$. In other words, the toric data ${\cal{G}}_2$ contains the columns of ${\cal{G}}_1$ and will have the form:
\begin{equation}
{\cal{G}}_2 = 
\left(
\begin{array}{ccccccccc}
 p_1 & p_2 & p_3 & \ldots & p_{\alpha } & p_{\alpha+1} & p_{\alpha+2} & \ldots & p_{\beta}  \\ \hline
 1 & 1 & 1 & \ldots & 1 & \ldots & \ldots & \ldots & 1 \\
 n_{21} & n_{22} & n_{23} & \ldots & n_{2 \alpha } & \ldots & \ldots & \ldots & n_{2 \beta } \\
 n_{31} & n_{32} & n_{33} & \ldots & n_{3 \alpha } & \ldots & \ldots & \ldots & n_{3 \beta } 
\end{array}
\right)~.
\label{Toric-data-G2}
\end{equation}
Note that we have labeled the columns as $p_1, p_2, ..., p_{\beta}$, which are the matter fields in Witten’s linear $\sigma$-model \cite{Feng:2000mi}.
The charge matrix corresponding to ${\cal{G}}_2$ is $Q_2$ satisfying $Q_2.{\cal{G}}_2^t=0$. This charge matrix $Q_2$ consists of $F$-term and $D$-term charge matrices \cite{Feng:2000mi}, $Q_F^2$ and $Q_D^2$, which are already known. Suppose there are $x$ number of rows in $Q_F^2$ and $y$ number of rows in $Q_D^2$. Thus, $Q_2$ will be given as:
\begin{equation}
Q_2 = 
\left[
\begin{array}{c}
 \left(Q_F^2\right)_{x \times \beta } \\ \hline
 \left(Q_D^2\right)_{y \times \beta }
\end{array}
\right] = 
\left(
\begin{array}{ccccc|c}
 p_1 & p_2 & p_3 & \ldots & p_{\beta } &  \\ \hline
 a_{11} & a_{12} & a_{13} & \ldots & a_{1 \beta } & 0 \\
 a_{21} & a_{22} & a_{23} & \ldots & a_{2 \beta } & 0 \\
 \vdots & \vdots & \vdots & \ldots & \vdots & \vdots \\
 a_{x1} & a_{x2} & a_{x3} & \ldots & a_{x \beta} & 0 \\ \hline
 b_{11} & b_{12} & b_{13} & \ldots & b_{1 \beta } & \zeta _1 \\
 b_{21} & b_{22} & b_{23} & \ldots & b_{2 \beta } & \zeta _2 \\
\vdots & \vdots & \vdots & \ldots & \vdots & \vdots \\
 b_{y1} & b_{y2} & b_{y3} & \ldots & b_{3 \beta} & \zeta _y
\end{array}
\right) ~.
\label{Q2-G2-CY2}
\end{equation}
Note that because of the relation $Q_2.{\cal{G}}_2^t=0$, the number of columns in $Q_2$ and hence $Q_F^2$ and $Q_D^2$ will be $\beta$. The $a$'s and $b$'s are all integers and are the elements of $Q_F^2$ and $Q_D^2$ respectively. We have also used the convention of \cite{Feng:2000mi} of introducing the last column $\left(0, 0, \ldots, 0, \zeta _1, \zeta _2, \ldots, \zeta _y  \right)$ to specify that the first set of rows are $F$-terms (and hence 0) and the second set of rows are $D$-terms (and hence resolved by the FI-parameters $\zeta _1, \zeta _2, \ldots, \zeta _y$).
 
In the partial resolution approach, we try to get the toric diagram of $CY_3^1$ by removing the points from the toric diagram of $CY_3^2$. This is equivalent to removing the extra columns from ${\cal{G}}_2$ to get the toric data ${\cal{G}}_1$. Thus we have to remove the set of columns $\{p_{\alpha+1}, p_{\alpha+2}, \ldots, p_{\beta} \}$ from the toric data ${\cal{G}}_2$ (\ref{Toric-data-G2}) to get the toric data ${\cal{G}}_1$ (\ref{Toric-data-G1}).

The next step is to determine the charge matrix $Q_1$ consisting of $Q_F^1$ and $Q_D^1$ charge matrices, so as to find the quiver gauge theory corresponding to $CY_3^1$. For this, we first write a row of $Q_1$ (say $r$) as a linear combination of rows of $Q_2$:
\begin{equation}
r = l_1 R_1 + l_2 R_2 + \ldots + l_x R_x + l_{x+1} R_{x+1} + l_{x+2} R_{x+2} + \ldots + l_{x+y} R_{x+y} ~,
\label{lin-comb-rows-Q2}
\end{equation}
where, $R_1, R_2, \ldots, R_x$ are the rows of $Q_F^2$ (\ref{Q2-G2-CY2}) and $R_{x+1}, R_{x+2}, \ldots, R_{x+y}$ are the rows of $Q_D^2$ (\ref{Q2-G2-CY2}). Substituting the rows from (\ref{Q2-G2-CY2}) into eq. (\ref{lin-comb-rows-Q2}), we will get a row $r$ of $Q_1$ in terms of rows of $Q_2$. Now, in order to get the toric data ${\cal{G}}_1$ from ${\cal{G}}_2$, we had to remove the set of columns $\{p_{\alpha+1}, p_{\alpha+2}, \ldots, p_{\beta} \}$. Thus, we must also remove these columns from the row $r$ (\ref{lin-comb-rows-Q2}). Equating the entries of the columns $\{p_{\alpha+1}, p_{\alpha+2}, \ldots, p_{\beta} \}$ in the row $r$ to 0, we get the following constraints respectively:
\begin{eqnarray}
\sum_{i=1}^{x} l_i a_{i, \alpha+1} + \sum_{i=1}^{y} l_{x+i} b_{i, \alpha+1} & = & 0 \nonumber \\
\sum_{i=1}^{x} l_i a_{i, \alpha+2} + \sum_{i=1}^{y} l_{x+i} b_{i, \alpha+2} & = & 0 \nonumber \\ \vdots \nonumber \\
\sum_{i=1}^{x} l_i a_{i \beta} + \sum_{i=1}^{y} l_{x+i} b_{i \beta} & = & 0 ~.
\label{constraints-Q2-TO-Q1}
\end{eqnarray}
From these constraints, we solve for the $l_1, l_2, ..., l_{x+y}$. These equations in $(x+y)$ number of $l_i$ variables can be solved into $m_z$ independent variables ($z \leq x+y$). Thus, the eq. (\ref{lin-comb-rows-Q2}), in terms of $m_j$ variables can be written as:
\begin{equation}
r = m_1 R_1^{'} + m_2 R_2^{'} + m_3 R_3^{'} + \ldots + m_z R_z^{'} ~,
\label{reduced-lin-comb-rows-Q2}
\end{equation}
where, each of the rows $R_1^{'}, R_2^{'}, \ldots, R_z^{'}$ is a known linear combination of the rows $R_1, R_2, \ldots, R_{x+y}$ of (\ref{Q2-G2-CY2}). Thus the charge matrix $Q_1$ is spanned by $z$ number of rows $R_1^{'}, R_2^{'}, \ldots, R_z^{'}$ and thus $Q_1$ can be fixed. However, just knowing $Q_1$ as whole will not solve the issue. We must also know which rows of $Q_1$ are the $Q_F^1$ rows and which are the $Q_D^1$ rows.

 To get the specific rows for $Q_F^1$ and $Q_D^1$, we must also keep track of the FI-parameters given in the last column of the $Q_2$ charge matrix (\ref{Q2-G2-CY2}). A combination of rows among the $Q_F^2$ rows will keep the last column of FI-parameters as 0. Thus any of the row $R_1^{'}, R_2^{'}, \ldots, R_z^{'}$ of $Q_1$ which is obtained as a result of combination of rows of $Q_F^2$, will continue to be the $F$-term row. Similarly, a row of $Q_1$ given as a combination of rows among $Q_D^2$, will give a non-zero FI-parameter and thus will continue to be a $D$-term row. However, any row of $Q_1$ which is formed by the combination between the $Q_F^2$ and $Q_D^2$ rows, will set a non-zero FI-parameter and such a row will become a $D$-term row. Suppose out of $z$ number of rows, $z_1$ correspond to a zero FI-parameter and $z_2$ number of rows correspond to non-zero FI-parameter (with $z=z_1 + z_2$). Thus the eq. (\ref{reduced-lin-comb-rows-Q2}) can be written as:
\begin{equation}
r = \left( m_1 R_1^{'} + m_2 R_2^{'} + \ldots + m_{z_1} R_{z_1}^{'} \right) + \left( m_{z_1+1} R_{z_1+1}^{'} + m_{z_2} R_{z_2}^{'} + \ldots m_{z_1+z_2} R_{z_1+z_2}^{'} \right) ~.
\label{final-reduced-lin-comb-rows-Q2}
\end{equation}
Thus, the rows $R_1^{'}, R_2^{'}, \ldots, R_{z_1}^{'}$ will form the $Q_F^1$ and the rows $R_{z_1+1}^{'}, R_{z_1+2}^{'}, \ldots, R_{z_1+z_2}^{'}$ will form the $Q_D^1$. Moreover, in writing the final form for $Q_1$, we must also remove the columns $\{p_{\alpha+1}, p_{\alpha+2}, \ldots, p_{\beta} \}$. This is because we have removed these columns in order to get ${\cal{G}}_1$ and we have already set the corresponding entries in $Q_1$ to 0 (\ref{constraints-Q2-TO-Q1}). Thus, the final expression for $Q_1$ can be given as:
\begin{equation}
Q_1 = \left[
\begin{array}{c}
 \left(Q_F^1\right)_{z_1 \times \alpha } \\ \hline
 \left(Q_D^1\right)_{z_2 \times \alpha }
\end{array}
\right] = 
\left(
\begin{array}{c}
 \left\{R_1^{'}\right\} \\
 \left\{R_2^{'}\right\} \\ 
\vdots \\
 \left\{R_{z_1}^{'}\right\} \\ \hline
 \left\{R_{z_1+1}^{'}\right\} \\
 \left\{R_{z_1+2}^{'}\right\} \\
\vdots \\
 \left\{R_{z_1+z_2}^{'}\right\}
\end{array}
\right)
\end{equation}
Note that this matrix $Q_1$ corresponding to $\toricG_1$, obtained as a result of partial resolution of $\toricG_2$, is sometimes also called as the reduced charge matrix. We will use this term frequently in this paper.

Once we have obtained $Q_F^1$ and $Q_D^1$, we can simply invert the steps of forward algorithm to get the quiver gauge theory \cite{Feng:2000mi} corresponding to $CY_3^1$. This is the procedure of partial resolution to get the quiver gauge theory for a toric Calabi-Yau by embedding it into a bigger toric diagram of another toric Calabi-Yau whose quiver gauge theory is already known. We have mentioned earlier that the toric diagram of any arbitrary toric $CY_3$, can be embedded into the toric diagram of $\left(\BZ_m \times \BZ_n\right)$ abelian orbifolds of $\BC^3$, for some suitable values of $m$ and $n$. Further, the quiver gauge theories for the abelian orbifolds of $\BC^3$ can be easily constructed from the structure of the orbifolding group $\left(\BZ_m \times \BZ_n\right)$ and are known in the literature. Thus partial resolution method, in principle, should be able to compute the quiver gauge theory for any given toric diagram. Partial resolutions of $\BC^3 /\left(\BZ_2 \times \BZ_2 \right)$ to get the conifold and suspended pinch point (SPP) theories have been investigated in \cite{Greene:1997uf, Muto:1997pq}. Partial resolution of $\BC^3 /\left(\BZ_3 \times \BZ_3 \right)$ was performed in \cite{Feng:2000mi} to get the theories corresponding to Fano twofolds. 
\subsection{Scheme of the paper}
Unfortunately, in the case of the abelian orbifolds of $\BC^4$, the quiver Chern-Simons theory is not known for a general $\left(\BZ_{n_1} \times \BZ_{n_2} \times \BZ_{n_3} \right)$ orbifold of $\BC^4$. Thus obtaining the quiver Chern-Simons theory of an arbitrary toric $CY_4$ by partial resolution method is not possible. However, we can still apply partial resolution approach, if we know that the toric diagram of a toric $CY_4$ can be embedded into the toric diagram of any bigger toric $CY_4$ (not necessarily an orbifold of $\BC^4$), whose quiver Chern-Simons theory is known. There are 18 toric Fano threefolds in literature and taking complex cone over them will give 18 toric $CY_4$. The quiver Chern-Simons theories corresponding to all these 18 toric $CY_4$ is known \cite{Davey:2011mz, Dwivedi:2011zm}. In this work, we first find if any of these 18 Fano threefolds can by embedded into any bigger Fano threefold. If such an embedding is possible, we carry out the partial resolution method to obtain the reduced charge matrices, using the steps given earlier. Once we have obtained the reduced charge matrix for a Fano threefold, we apply inverse algorithm to see if it is possible to extract a quiver diagram. A quiver diagram consists of nodes representing the $U(N)$ gauge groups. An arrow connecting node-$a$ to node-$b$ is a bifundamental field transforming under the fundamental representation of $U(N)_a$ and the anti-fundamental representation of $U(N)_b$. There can also be an arrow which start and end at same node-$c$, which represents an adjoint field transforming under the adjoint representation of $U(N)_c$. The matter content given by the quiver diagram can be encoded into a matrix, called as the incidence matrix \cite{Feng:2000mi} which is given as:
\begin{equation}
\left(d\right)_{G \times E} = 
\left(
\begin{array}{ccccc}
 X_1 & X_2 & X_3 & \ldots & X_E \\ \hline
 d_{11} & d_{12} & d_{13} & \ldots & d_{1E} \\
 d_{21} & d_{22} & d_{23} & \ldots & d_{2E} \\
\vdots & \vdots & \vdots & \vdots & \vdots \\
 d_{G1} & d_{G2} & d_{G3} & \ldots & d_{GE}
\end{array}
\right)~.
\label{incidence-matrix}
\end{equation} 
Here $G$ labels the gauge groups or the nodes in the quiver diagram and $X_i$'s label the fields or the arrows in the quiver diagram. We can set a convention to fix the entries of the $d$-matrix by looking at a quiver diagram. Similar to the convention used in \cite{Feng:2000mi}, we use the following convention in this paper:
\begin{equation}
\left(d \right)_{\alpha, X_{\beta}} = {\delta}_{Tail(X_{\beta}), \alpha} - \delta_{Head(X_{\beta}), \alpha }\nonumber ~.
\label{quiverfromd}
\end{equation}
This convention means that if there is an arrow $X_{\beta}$ in the quiver diagram which  starts at the node-$\alpha$ but ends at any other node, the element of $d$ corresponding to node-$\alpha$ and field-$X_{\beta}$ will read -1. Similarly if the arrow $X_{\beta}$ starts at some other node and ends on node-$\alpha$, the corresponding element of $d$-matrix will be 1. If there is an adjoint field represented by arrow $X_{\beta}$ starting and ending on the same node-$\alpha$, the corresponding entry of $d$ will be 0. It is not difficult to see that in the matrix $d$ given in (\ref{incidence-matrix}), an element can only take values 1, -1 or 0. With this restriction in mind, we apply the inverse algorithm and extract $d$ to see what kind of quiver diagrams we can construct. It should also be mentioned that these quiver diagrams should have equal number of incoming and outgoing arrows at each node.

It is interesting to mention that we tried removing various possible points in the toric diagrams of each of the 18 Fano threefolds and were able to detect the following embeddings (some of these embeddings are already known, in the references given):
\begin{itemize}
	\item Fano $\BP^3$ is embedded inside Fano $\Bc$ \cite{Dwivedi:2012qa} , Fano $\Bb$ \cite{Dwivedi:2012qa} and Fano $\Ce$.
	\item Fano $\Ba$ is embedded inside Fano $\Cb$. 
	\item Fano $\Bd$ is embedded inside Fano $\Db$ \cite{Phukon:2011hp} and Fano $\Cd$ \cite{Phukon:2011hp}. 
	\item Fano $\Ca$ is embedded inside Fano $\Eb$. 
	\item Fano $\Cc$ is embedded inside Fano $\Ec$ \cite{Phukon:2011hp} and Fano $\Fa$. 
	\item Fano $\Db$ is embedded inside Fano $\Ed$. 
	\item Fano $\Ea$ is embedded inside Fano $\Fb$ \cite{Phukon:2011hp}. 
\end{itemize}
In the following sections, we will discuss these embeddings in detail. We will obtain the reduced charge matrices via partial resolution and extract the quiver diagrams, if possible.

\section{Embedding of Fano $\BP^3$ inside other Fano threefolds}
\label{sec3}
In this section, we will discuss about the possible embeddings of Fano $\BP^3$ inside the remaining 17 Fano threefolds. We tried all the possibilities of removing points from the toric diagrams of these 17 Fano threefolds to get the toric diagram of Fano $\BP^3$. We find that Fano $\BP^3$ can be embedded inside Fano $\Ce$, Fano $\Bc$ and Fano $\Bb$. The embedding of Fano $\BP^3$ inside Fano $\Bc$ and Fano $\Bb$ has already been done in \cite{Dwivedi:2012qa}. We will first review this result in the following two subsections as a warmup, and then discuss about the embedding of Fano $\BP^3$ inside Fano $\Ce$. Since the quiver Chern-Simons theories for all the 18 Fano threefolds are known, we can also perform the partial resolution of Fano $\Bc$, Fano $\Bb$ and Fano $\Ce$ to obtain the reduced charge matrix for Fano $\BP^3$. From the reduced charge matrix, we can obtain the possible quiver diagrams using the inverse algorithm method. We have shown by an example in section 3.3, on how to extract the possible quiver diagrams from a reduced charge matrix.  If we can find a quiver gauge theory for Fano $\BP^3$, whose quiver diagram is different from what is already known in the literature \cite{Dwivedi:2011zm}, we can say that we have obtained a new toric duality. The partial resolution of Fano $\Bc$ and Fano $\Bb$ was given in \cite{Dwivedi:2012qa}. We will first review them and then perform the partial resolution on Fano $\Ce$. 
\subsection{Embedding of Fano $\BP^3$ inside Fano $\Bc$}
\begin{figure}[tbp]
	\centering
		\includegraphics[width=1.05\textwidth]{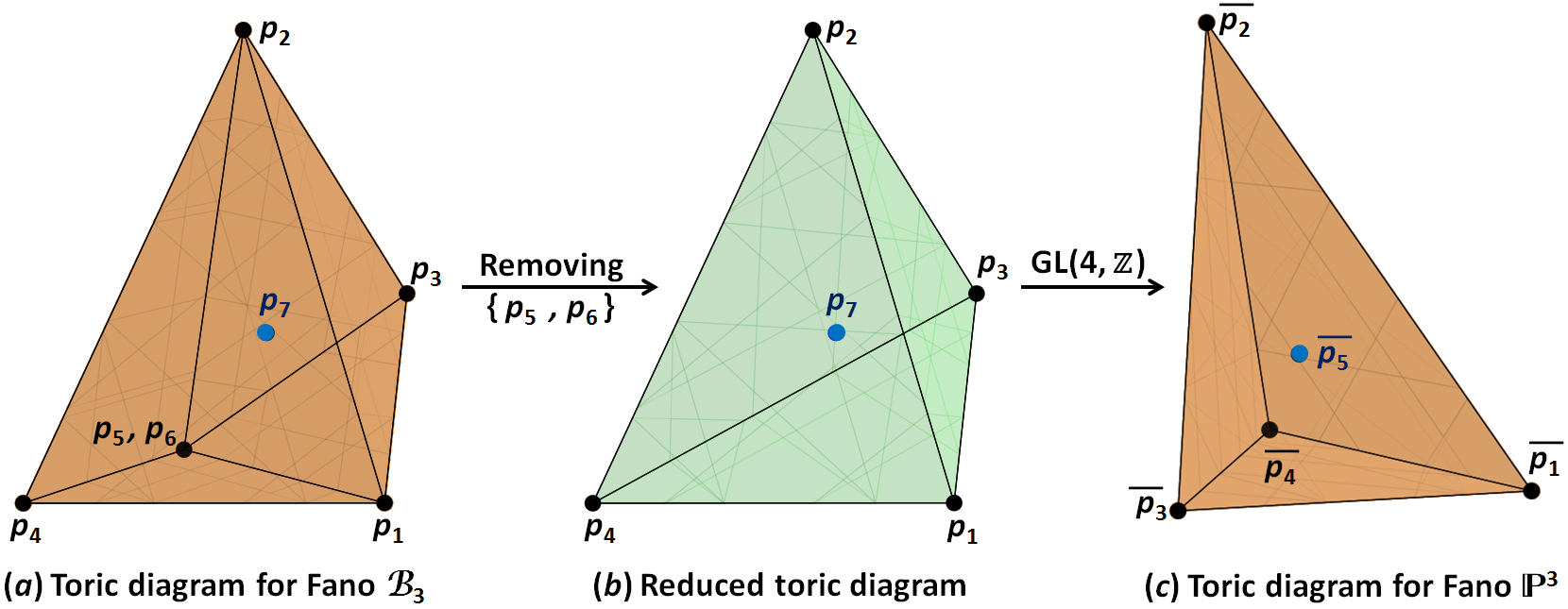}
	\caption{Figure ($a$) shows the toric diagram corresponding to the toric data (\ref{G-B3}) along with The GLSM fields $p_i$. Note that $p_5$ and $p_6$ correspond to the same point implying a multiplicity 2. Figure ($b$) is the reduced toric diagram obtained by removing $p_5$ and $p_6$. This is equivalent to the toric diagram for Fano $\BP^3$ in figure ($c$) whose toric data is given in (\ref{G-P3}).}
	\label{ToricB3toP3}
\end{figure}

The toric diagram of Fano $\Bc$ can be encoded into the following toric data:
\begin{equation}
{\cal{G}}_{\Bc} = \left(
\begin{array}{ccccccc}
p_1 & p_2 & p_3 & p_4 & p_5 & p_6 & p_7 \\ \hline
 1 & 1 & 1 & 1 & 1 & 1 & 1 \\
 1 & -1 & 0 & 0 & 0 & 0 & 0 \\
 0 & 0 & 1 & -1 & 0 & 0 & 0 \\
 0 & 1 & 0 & -1 & -1 & -1 & 0
\end{array}
\right)~.
\label{G-B3}
\end{equation}
The vectors forming the column of this toric data are four dimensional but the tip of all these vectors lie on the same hyperplane. Thus the first entries of each column is 1. The remaining three components of each column vector can be represented as a point in $\BZ^3$ lattice. The convex hull of all these points is a convex lattice polygon and is called the toric diagram. The toric diagram of Fano $\Bc$ encoded by the toric data (\ref{G-B3}) is shown in figure \ref{ToricB3toP3}($a$). The point ($0,0,0$) is an internal point in the toric diagram and is shown in blue. In fact, the toric diagrams of all the Fano threefolds have  ($0,0,0$) as an internal point \cite{Davey:2011mz}. From the toric data, we see that each column is associated with a GLSM field $p_i$ which are matter fields in Witten’s linear $\sigma$-model. Note that it may be possible for different $p_i$ fields to be associated with the same point in the toric diagram. For example, the fields $p_5$ and $p_6$ correspond to the same point ($0,0,-1$) of the toric diagram as shown in figure \ref{ToricB3toP3}($a$). Thus the point ($0,0,-1$) has multiplicity 2, all other points have multiplicities 1. As far as toric diagram is concerned, the multiplicities do not have any role. However they have a crucial role in determining the corresponding quiver gauge theory and hence we will keep track of the multiplicities.  

The toric data for Fano $\BP^3$ is given as:
 \begin{equation}
{\cal{G}}_{\BP^3} = \left(
\begin{array}{ccccc}
\overline{p_1} & \overline{p_2} & \overline{p_3} & \overline{p_4} & \overline{p_5} \\ \hline
 1 & 1 & 1 & 1 & 1 \\
 1 & -1 & 0 & 0 & 0 \\
 0 & 1 & -1 & 0 & 0 \\
 0 & 0 & 1 & -1 & 0
\end{array}
\right)~,
\label{G-P3}
\end{equation}
and is shown in figure \ref{ToricB3toP3}($c$), where we have written the GLSM fields as $\overline{p_i}$ to differentiate them from the GLSM fields of Fano $\Bc$. From the next section onwards, we will not differentiate them and the GLSM fields in a toric data will be only represented by $p_i$.
As shown in \cite{Dwivedi:2012qa}, if we remove the set of points $\left\{p_5, p_6\right\}$ from the toric data of Fano $\Bc$ (\ref{G-B3}), we get a reduced toric data shown in figure \ref{ToricB3toP3}($b$). This is equivalent to the toric data of $\BP^3$ because toric data of $\BP^3$ can be obtained by acting a $GL(4,\BZ)$ transformation on the reduced toric data as given below:
\begin{equation}
{\cal{G}}_{\BP^3} = 
\left(
\begin{array}{cccc}
1 & 0 & 0 & 0 \\
 0 & 1 & 0 & 0 \\
 0 & 0 & -1 & 1 \\
 0 & 0 & 1 & 0
\end{array}
\right).
\left(
\begin{array}{ccccc}
p_1 & p_2 & p_3 & p_4 & p_7 \\ \hline
 1 & 1 & 1 & 1 & 1 \\
 1 & -1 & 0 & 0 & 0 \\
 0 & 0 & 1 & -1 & 0 \\
 0 & 1 & 0 & -1 & 0
\end{array}
\right)
\end{equation} 
We have also shown this in the figure \ref{ToricB3toP3}. Thus Fano $\BP^3$ is embedded inside Fano $\Bc$. Next, we will perform the partial resolution of Fano $\Bc$ \cite{Dwivedi:2012qa}. For this, we will need the information about the quiver Chern-Simons theory corresponding to Fano $\Bc$, which is given in \cite{Dwivedi:2011zm}. The charge matrix corresponding to the quiver Chern-Simons theory for Fano $\Bc$ is given below \cite{Dwivedi:2011zm}:
\begin{equation}
Q_{\Bc} = \left(
\begin{array}{c}
 Q_F^{\Bc} \\ \hline
 Q_D^{\Bc}
\end{array}
\right)
=
\left(
\begin{array}{ccccccc}
 1 & 1 & 3 & 3 & -1 & -1 & -6 \\
 1 & 1 & 1 & 1 & 0 & 0 & -4 \\ \hline
0 & 0 & 2 & 2 & -2 & 0 & -2
\end{array}
\right)~.
\label{Q-B3}
\end{equation}
As discussed in section \ref{sec2}, we can write a row ($r$) of the charge matrix $Q_{\BP^3}$ for Fano $\BP^3$ as a linear combination of the rows ($R_i$) of charge matrix for Fano $\Bc$ given in eq. (\ref{Q-B3}) as:
\begin{eqnarray}
r & = & a_1 R_1 + a_2 R_2 + a_3 R_3 \nonumber \\
& = & (a_1+a_2, a_1+a_2, 3a_1+a_2+2a_3, 3a_1+a_2+2a_3,-a_1-2a_3,-a_1, -6a_1-4a_2-2a_3) \nonumber ~.
\end{eqnarray}
Since we deleted the points $\{p_5, p_6\}$ in order to get the toric data of Fano $\BP^3$  from Fano $\Bc$, we must also set the corresponding columns 5, 6 in $r$ to 0. Thus, we get:
\begin{equation}
-a_1-2a_3 = 0, \quad -a_1 = 0 \quad \Rightarrow \quad a_1 = a_3 = 0 ~.
\end{equation}  
Substituting it back in $r$, we get:
\begin{eqnarray}
r  =  a_2 R_2 = a_2(1, 1, 1, 1, 0, 0, -4) ~.
\end{eqnarray}
Thus, $Q_{\BP^3}$ will be spanned by just one row. Since this row was a $Q_F$ row in (\ref{Q-B3}), it will continue to be an $F$-term row. Thus, the charge matrix $Q_{\BP^3}$ (after deleting the columns 5 and 6) will be given as: 
\begin{equation}
Q_{\BP^3} = (Q_F^{\BP^3}) = (1,1,1,1,-4)~,
\label{Q-reduced-P3}
\end{equation}
which is the charge matrix of Fano $\BP^3$. This matches with the charge matrix for Fano $\BP^3$ theory known in the literature \cite{Dwivedi:2011zm} and we will get the same quiver diagram for Fano $\BP^3$ as given in \cite{Dwivedi:2011zm}.
\subsection{Embedding of Fano $\BP^3$ inside Fano $\Bb$}
\begin{figure}[tbp]
	\centering
		\includegraphics[width=0.9\textwidth]{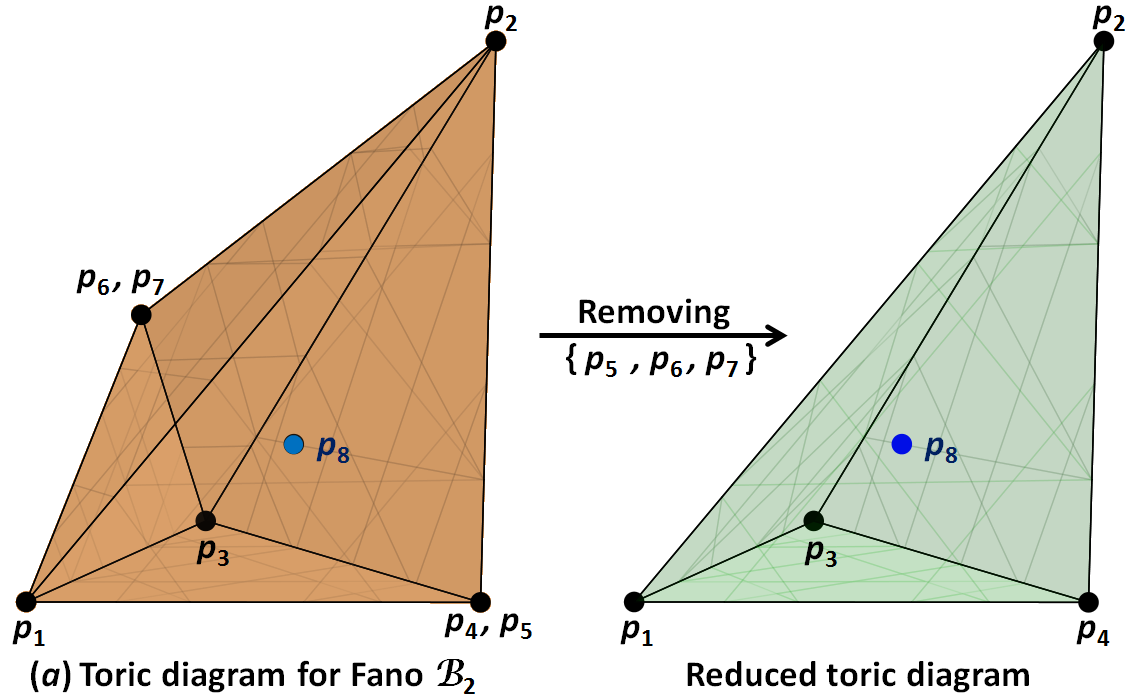}
	\caption{Figure ($a$) shows the toric diagram for Fano $\Bb$ corresponding to the toric data (\ref{G-B2}). Removing $\left\{p_5, p_6, p_7\right\}$ gives a reduced toric diagram which is equivalent to toric diagram for Fano $\BP^3$ of figure \ref{ToricB3toP3}($c$).}
	\label{ToricB2toP3}
\end{figure}

The toric data of Fano $\Bb$ is given below:
\begin{equation}
{\cal{G}}_{\Bb} = \left(
\begin{array}{cccccccc}
p_1 & p_2 & p_3 & p_4 & p_5 & p_6 & p_7 & p_8 \\ \hline
 1 & 1 & 1 & 1 & 1 & 1 & 1 & 1 \\
 1 & -1 & 0 & 0 & 0 & 0 & 0 & 0 \\
 0 & 1 & -1 & 0 & 0 & 0 & 0 & 0 \\
 0 & 0 & 1 & -1 & -1 & 1 & 1 & 0
\end{array}
\right)~.
\label{G-B2}
\end{equation}
The toric diagram is shown in figure \ref{ToricB2toP3}($a$).
If we remove the set of points $\left\{p_5, p_6, p_7\right\}$ \cite{Dwivedi:2012qa} from the toric data of Fano $\Bb$ (\ref{G-B2}), we get a reduced toric data which is related by a $GL(4,\BZ)$ transformation  to the toric data of Fano $\BP^3$ (\ref{G-P3}):
\begin{equation}
{\cal{G}}_{\BP^3} = 
\left(
\begin{array}{cccc}
1 & 0 & 0 & 0 \\
0 & 1 & 0 & 0 \\
0 & 0 & 1 & 0 \\
0 & 0 & 0 & 1
\end{array}
\right).
\left(
\begin{array}{ccccc}
p_1 & p_2 & p_3 & p_4 & p_8 \\ \hline
 1 & 1 & 1 & 1 & 1 \\
 1 & -1 & 0 & 0 & 0 \\
 0 & 1 & -1 & 0 & 0 \\
 0 & 0 & 1 & -1 & 0
\end{array}
\right)
\end{equation} 
Thus, Fano $\BP^3$ is embedded inside Fano $\Bb$. Performing the partial resolution of Fano $\Bb$ \cite{Dwivedi:2012qa}, we obtain the same reduced charge matrix for Fano $\BP^3$ given in (\ref{Q-reduced-P3}). Hence we will get the same quiver diagram for Fano $\BP^3$ shown in \cite{Dwivedi:2011zm}.  

\subsection{Embedding of Fano $\BP^3$ inside Fano $\Ce$}
Now we will discuss about the embedding of Fano $\BP^3$ inside Fano $\Ce$. It is interesting to note that Fano $\Ce$ itself has two toric dual quiver gauge theories, known as phase-I and phase-II of Fano $\Ce$ as shown in figure \ref{Phases-C5}, which have been discussed in \cite{Davey:2011mz, Dwivedi:2014awa}. Both of these phases have the same toric data as that of Fano $\Ce$, but with different multiplicity of one of the point. This means that the number of columns in the toric data for phase-I and phase-II of Fano $\Ce$ are different, though both of them represent the same toric diagram of complex cone over Fano $\Ce$. If the number of columns in toric data are different, the number of $p_i$ fields we have to remove in order to show the embedding of Fano $\BP^3$, will also be different. This will affect the partial resolution approach, and in principle, may lead to different reduced charge matrices. Thus, we will do the partial resolution of both the phases of Fano $\Ce$ to get the reduced charge matrices for Fano $\BP^3$.

\subsubsection{Partial resolution of Phase-I of Fano $\Ce$}
\begin{figure}
	\centering
		\includegraphics[width=0.75\textwidth]{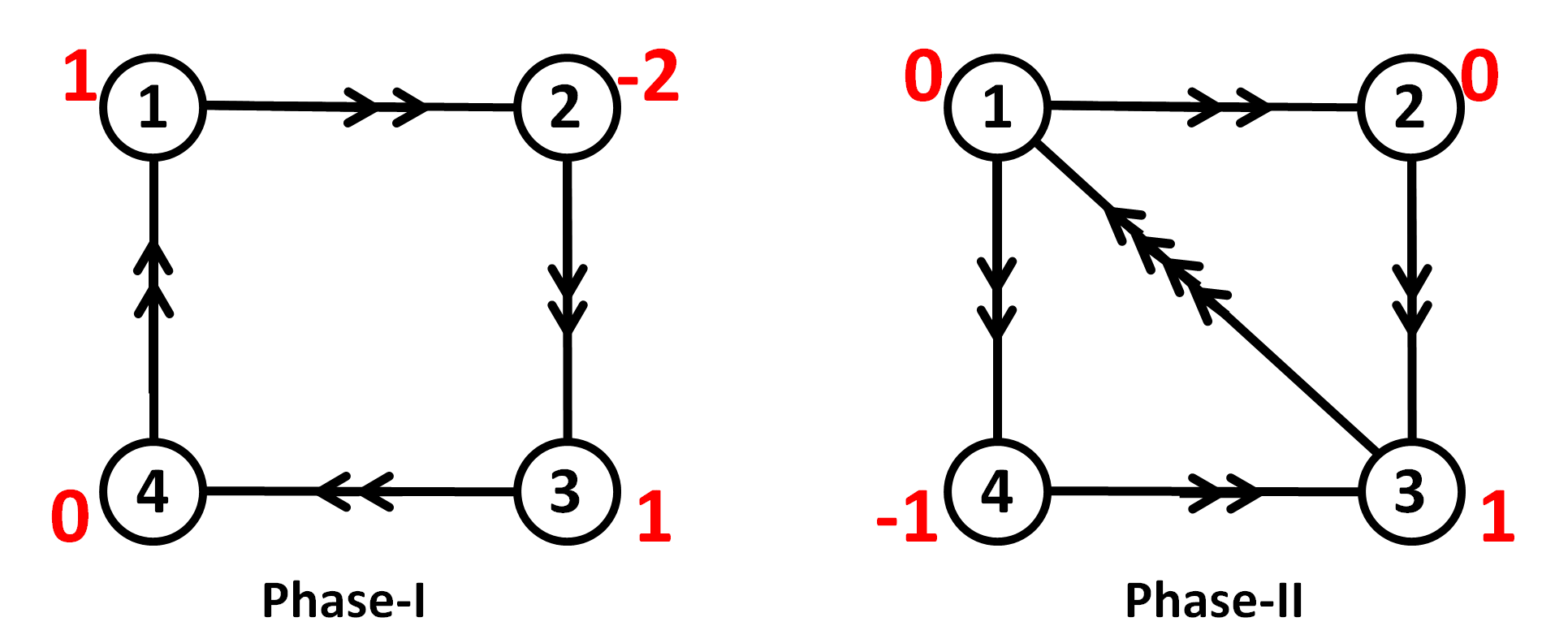}
	\caption{Quiver diagram for Phase-I and Phase-II of Fano $\Ce$. The Chern-Simons levels are given in red adjacent to the nodes.}
	\label{Phases-C5}
\end{figure}

The quiver gauge theory for phase-I of Fano $\Ce$ has been discussed in \cite{Davey:2011mz}. The quiver diagram for this theory is given in figure \ref{Phases-C5}.
Further the toric data (${\cal{G}}$) of the $CY_4$ which is the complex cone over Fano $\Ce$ corresponding to phase-I is given as \cite{Davey:2011mz}:
\begin{equation}
{\cal{G}}^{Phase-I}_{\Ce} = \left(
\begin{array}{cccccccc}
 p_1 & p_2 & p_3 & p_4 & p_5 & p_6 & p_7 & p_8  \\ \hline
 1 & 1 & 1 & 1 & 1 & 1 & 1 & 1 \\
 1 & -1 & 0 & 0 & 0 & 0 & 0 & 0 \\
 0 & 0 & 1 & -1 & 0 & 0 & 0 & 0 \\
 0 & 1 & 0 & -1 & 1 & -1 & 0 & 0
\end{array}
\right) ~,
\label{G-phase-I-C5}
\end{equation}
The corresponding toric diagram is shown in figure \ref{ToricphaseIC5toP3}($a$). We find that if we remove the set of columns $\{p_5,p_6,p_7\}$ or $\{p_5,p_6,p_8\}$, we get a reduced toric data which is $GL(4, \BZ)$ related to the toric data of Fano $\BP^3$ (\ref{G-P3}):
\begin{figure}[tbp]
	\centering
		\includegraphics[width=0.9\textwidth]{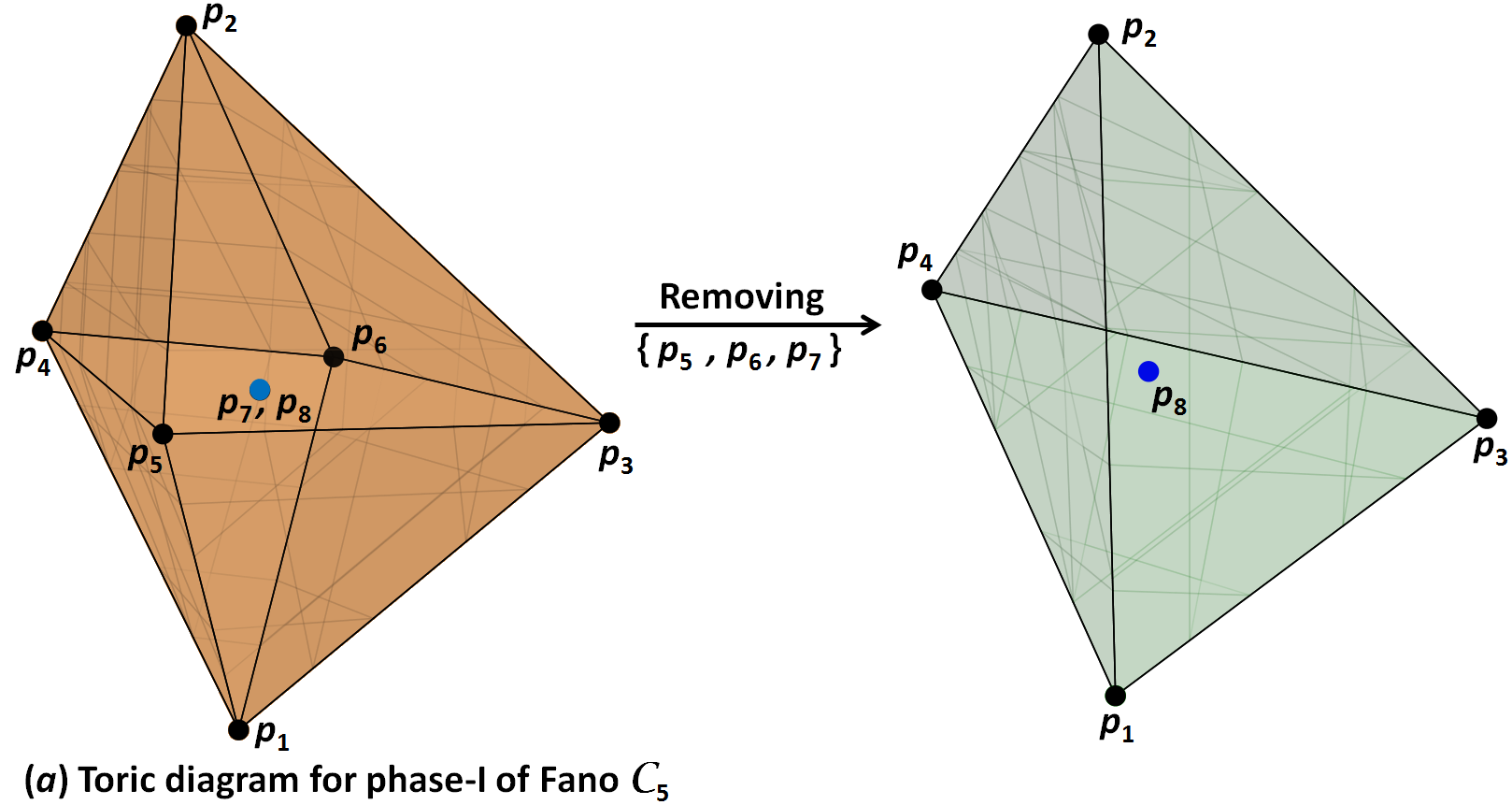}
	\caption{Figure ($a$) shows the toric diagram for phase-I of Fano $\Ce$. Removing either $\left\{p_5, p_6, p_7\right\}$ or  $\{p_5,p_6,p_8\}$ gives a toric diagram which is equivalent to toric diagram for Fano $\BP^3$ of figure \ref{ToricB3toP3}($c$).}
	\label{ToricphaseIC5toP3}
\end{figure}

\begin{eqnarray}
{\cal{G}}_{\BP^3} & = & \left(
\begin{array}{cccc}
 1 & 0 & 0 & 0 \\
 0 & 1 & 0 & 0 \\
 0 & 0 & -1 & 1 \\
 0 & 0 & 1 & 0
\end{array}
\right).\left(
\begin{array}{ccccc}
 p_1 & p_2 & p_3 & p_4 & p_8 \\ \hline
 1 & 1 & 1 & 1 & 1 \\
 1 & -1 & 0 & 0 & 0 \\
 0 & 0 & 1 & -1 & 0 \\
 0 & 1 & 0 & -1 & 0
\end{array}
\right)~, \nonumber \\ 
{\cal{G}}_{\BP^3} & = & \left(
\begin{array}{cccc}
 1 & 0 & 0 & 0 \\
 0 & 1 & 0 & 0 \\
 0 & 0 & -1 & 1 \\
 0 & 0 & 1 & 0
\end{array}
\right).\left(
\begin{array}{ccccc}
 p_1 & p_2 & p_3 & p_4 & p_7 \\ \hline
 1 & 1 & 1 & 1 & 1 \\
 1 & -1 & 0 & 0 & 0 \\
 0 & 0 & 1 & -1 & 0 \\
 0 & 1 & 0 & -1 & 0
\end{array}
\right) ~.
\label{G-from-phase-I-C5-to-P3}
\end{eqnarray}
Note that the $p_7$ and $p_8$ columns in ${\cal{G}}^{Phase-I}_{\Ce}$ are repeated columns. Thus removing either of $p_7$ or $p_8$ will not make any difference as far as toric datas are concerned. This is also evident from eq. (\ref{G-from-phase-I-C5-to-P3}). However, this may lead to different partial resolutions.

The next step is to find the reduced charge matrix (say $Q$) for Fano $\BP^3$. We start with the charge matrix for the quiver gauge theory for phase-I of Fano $\Ce$, which is given by \cite{Davey:2011mz}:
\begin{equation}
Q^{Phase-I}_{\Ce} =\left(
\begin{array}{c}
 Q_F \\ \hline
 Q_D
\end{array}
\right)=\left(
\begin{array}{cccccccc}
 1 & 1 & 0 & 0 & -1 & 0 & -1 & 0 \\
 0 & 0 & 1 & 1 & 0 & -1 & 0 & -1 \\ \hline
 0 & 0 & 0 & 0 & 1 & 1 & 0 & -2 \\
 0 & 0 & 0 & 0 & 0 & 0 & 1 & -1
\end{array}
\right) ~.
\label{Q-Phase-I-C5}
\end{equation}
Since we are looking for the possible embedding, a row ($r$) of the reduced charge matrix $Q$ for Fano $\BP^3$ can be written as linear combination of rows of the charge matrix of Fano $\Ce$, given in eq. (\ref{Q-Phase-I-C5}).
Thus, we can write,
\begin{equation}
r = a_1R_1 + a_2R_2 + a_3R_3 + a_4R_4 ~,
\end{equation}
where $R_i$'s denote the rows of the charge matrix of Fano $\Ce$. So, we have,
\begin{equation}
r = \left(
\begin{array}{cccccccc}
 a_1, & a_1, & a_2, & a_2, & -a_1+a_3, & -a_2+a_3, & -a_1+a_4, & -a_2-2 a_3-a_4
\end{array}
\right) ~.
\label{lin-comb-phase-I-C5}
\end{equation}
We have seen earlier, we have the option of either removing $\{p_5,p_6,p_7\}$ or $\{p_5,p_6,p_8\}$ from the toric diagram. Let us first remove $\{p_5,p_6,p_7\}$. In the partial resolution method, we must also set the corresponding columns in the row given in eq. (\ref{lin-comb-phase-I-C5}) and should remove it. Setting these columns to 0, we will get $-a_1+a_3=0$, $-a_2+a_3=0$ and $-a_1+a_4=0$ respectively. This gives $a_4 = a_3 = a_2 = a_1$. Thus, a row of the reduced charge matrix will be given as,
\begin{eqnarray}
r & = & a_1R_1 + a_1R_2 + a_1R_3 + a_1R_4 \nonumber \\ 
& = & a_1 (R_1 + R_2 + R_3 + R_4) 
\end{eqnarray}
Thus we find that the required $Q$ is spanned by just one row which is $R_1 + R_2 + R_3 + R_4 = (1, 1, 1, 1, 0, 0, 0, -4)$. Since this is a combination of the $Q_F$ and $Q_D$ rows as given in (\ref{Q-Phase-I-C5}), the row $R_1 + R_2 + R_3 + R_4$ will now correspond to a $D$-term and hence will be a $Q_D$ row in the reduced charge matrix. Hence, the reduced charge matrix for Fano $\BP^3$ is given as,
\begin{equation}
Q = Q_D = (1, 1, 1, 1, -4) ~,
\label{Q-new-P3}
\end{equation}
where we have also removed the columns 5, 6, 7. We see that in this case, we obtain $Q_F = 0$ (i.e. all the entries in $Q_F$ are 0) and so the total charge matrix ($Q$) contains only $Q_D$. This is a possible charge matrix for Fano $\BP^3$. This charge matrix is different from the existing charge matrix for Fano $\BP^3$ in the literature \cite{Dwivedi:2011zm} given in eq. (\ref{Q-reduced-P3}).
So our next step must be to continue with the charge matrix of eq. (\ref{Q-new-P3}), apply inverse algorithm and see if we can get a new quiver diagram. From the $Q_F$ matrix, we obtain matrix $T$ which is nullspace of $Q_F$ ($T.Q_F^t=0$) \cite{Feng:2000mi}. From $T$, we can obtain the dual cone matrix $K$ (see appendix of \cite{Feng:2000mi} on how to find the dual cone) such that $K.T \geq 0$, which means all the elements of matrix $K.T$ are non-negative. For our present case, we obtain these matrices as,
\begin{equation}
T = \left(
\begin{array}{ccccc}
 0 & 0 & 0 & 0 & 1 \\
 0 & 0 & 0 & 1 & 0 \\
 0 & 0 & 1 & 0 & 0 \\
 0 & 1 & 0 & 0 & 0 \\
 1 & 0 & 0 & 0 & 0 \\
\end{array}
\right)~; K = \left(
\begin{array}{ccccc}
 0 & 0 & 0 & 0 & 1 \\
 0 & 0 & 0 & 1 & 0 \\
 0 & 0 & 1 & 0 & 0 \\
 0 & 1 & 0 & 0 & 0 \\
 1 & 0 & 0 & 0 & 0 \\
\end{array}
\right)~.
\end{equation}
We define another matrix $P\equiv K.T$ which is called as the perfect matching matrix. All the entries of this matrix are non-negative. The rows of this matrix label the matter fields or the arrows $X_i$ of the quiver diagram and columns label the GLSM fields $p_i$. In other words this matrix encodes the GLSM charges of various arrows of the quiver diagram. Here we get this matrix as,
\begin{equation}
P = \left(
\begin{array}{c|ccccc}
& p_1 & p_2 & p_3 & p_4 & p_5 \\ \hline
X_1 & 1 & 0 & 0 & 0 & 0 \\
X_2 & 0 & 1 & 0 & 0 & 0 \\
X_3 & 0 & 0 & 1 & 0 & 0 \\
X_4 & 0 & 0 & 0 & 1 & 0 \\
X_5 & 0 & 0 & 0 & 0 & 1 \\
\end{array}
\right) ~.
\label{P-new-P3}
\end{equation}
Using $P$ and $Q_D$ matrices, we can obtain a matrix $\Delta_{(G-2) \times E}$ called as projected charge matrix and is given as $\Delta^t=P.Q_D^t$. Here $G$ indicates the number of nodes in the required quiver diagram and $E$ is the number of arrows in the quiver diagram. The matrix $\Delta$ is related to the incidence matrix $d$ explained in eq. (\ref{incidence-matrix}) and the Chern-Simons levels ($\{k_i \}$) of the theory by the relation,
\begin{equation}
\Delta_{ij} = k_{i+1}d_{ij} - k_{i}d_{(i+1)j} \quad, \quad i=(1,2,\ldots,G-2)~.
\label{Deltaij}
\end{equation}  
Here the matrix $d$ and Chern-Simons levels ($k_1, k_2, \dots, k_G$) are the unknowns along with the constraints that the elements of matrix $d$ can only be 1, -1 or 0 and the sum of Chern-Simons levels of all the nodes vanishes ($\sum_{i=1}^G k_i = 0$). Thus we can obtain all possible combinations of $d$ matrices and Chern-Simons levels. From $d$-matrix, we can draw the quiver diagram using the convention given in eq. (\ref{quiverfromd}). The superpotential $W$ of the quiver gauge theory can be constructed from $K$ matrix \cite{Feng:2000mi}.

\begin{figure}
	\centering
		\includegraphics[width=0.90\textwidth]{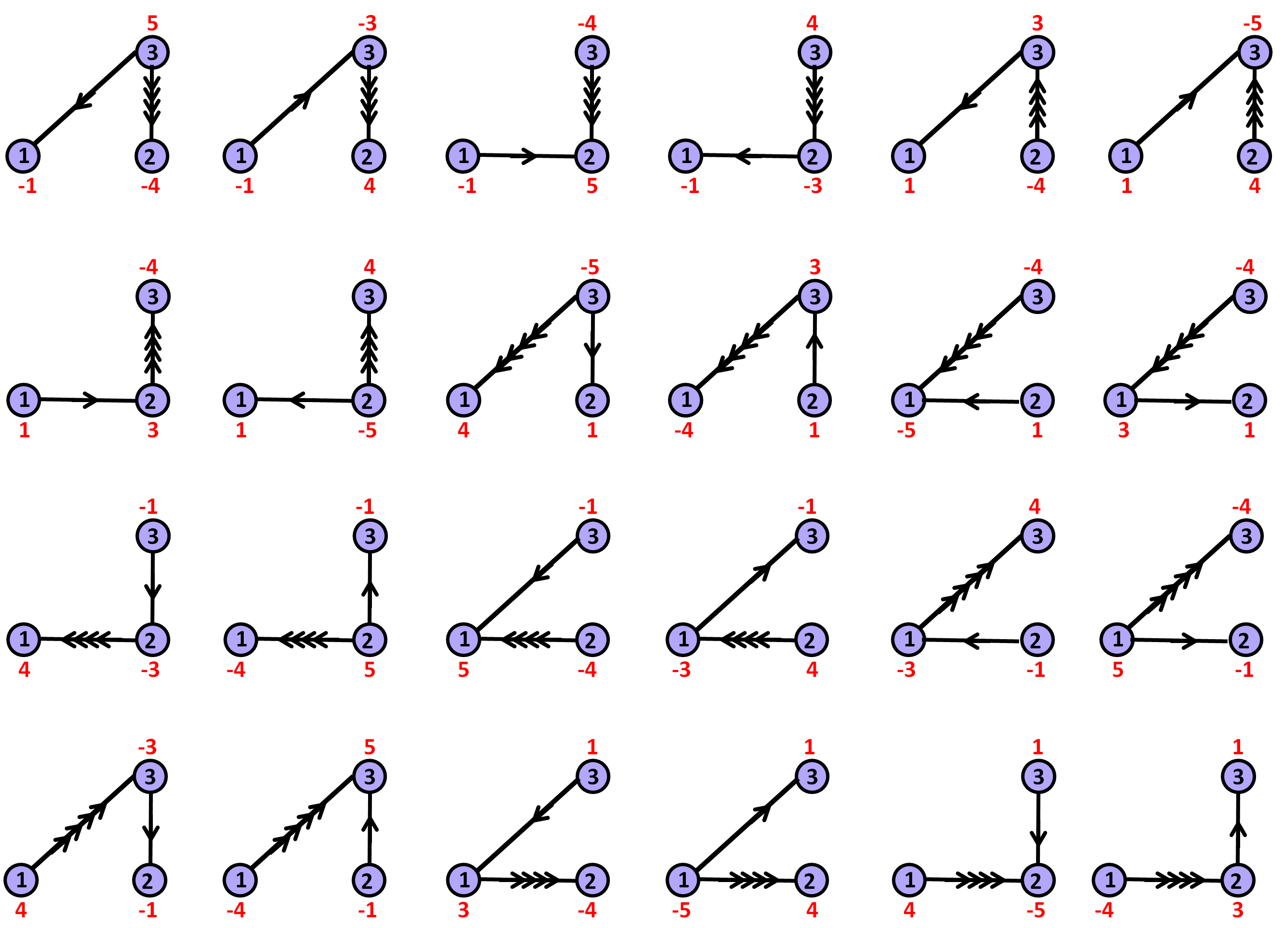}
	\caption{24 diagrams corresponding to the charge matrix (\ref{Q-new-P3}). These diagrams violate the restriction that a quiver gauge theory has equal number of incoming and outgoing arrows.}
	\label{Phase-I-C5-INTO-P3}
\end{figure}

For our present case, using $Q_D$ and $P$ matrices given in eq. (\ref{Q-new-P3}) and eq. (\ref{P-new-P3}) respectively, we obtain,
\begin{equation}
\Delta = \left(
\begin{array}{ccccc}
 1 & 1 & 1 & 1 & -4 \\
\end{array}
\right)~.
\end{equation}
From this $\Delta$, we obtain 24 different diagrams as shown in figure \ref{Phase-I-C5-INTO-P3}, but these diagrams are not the quiver diagrams because they do not have equal number of incoming and outgoing arrows at each node.

We further tried the second option of removing $\{p_5,p_6,p_8\}$ from the toric data of phase-I of Fano $\Ce$ and obtained the same charge matrix given in eq. (\ref{Q-new-P3}) which means we get the same diagrams as shown in figure \ref{Phase-I-C5-INTO-P3}.

\subsubsection{Partial resolution of Phase-II of Fano $\Ce$}
The quiver diagram for phase-II of Fano $\Ce$ \cite{Davey:2011mz} is given in figure \ref{Phases-C5}.
The toric data (${\cal{G}}$) of the complex cone over Fano $\Ce$ corresponding to phase-II is given as \cite{Davey:2011mz}:
\begin{equation}
{\cal{G}}^{Phase-II}_{\Ce} = \left(
\begin{array}{ccccccccc}
 p_1 & p_2 & p_3 & p_4 & p_5 & p_6 & p_7 & p_8 & p_9 \\ \hline
 1 & 1 & 1 & 1 & 1 & 1 & 1 & 1 & 1 \\
 1 & -1 & 0 & 0 & 0 & 0 & 0 & 0 & 0 \\
 0 & 0 & 1 & -1 & 0 & 0 & 0 & 0 & 0 \\
 0 & 1 & 0 & -1 & 1 & -1 & 0 & 0 & 0
\end{array}
\right) ~.
\label{G-phase-II-C5}
\end{equation}
The toric diagram drawn from toric data (\ref{G-phase-II-C5}) is same as shown in figure \ref{ToricphaseIC5toP3}(a) except that the internal point ($0,0,0$) now has multiplicity 3 because the GLSM fields $\{p_7,p_8,p_9\}$ correspond to the same point ($0,0,0$). We can clearly see that to get the toric diagram of Fano $\BP^3$ given by toric data (\ref{G-P3}), we have now 3 choices of removing points from toric data (\ref{G-phase-II-C5}): $\{p_5,p_6,p_7,p_8\}$, $\{p_5,p_6,p_7,p_9\}$, $\{p_5,p_6,p_8,p_9\}$. However we find that in all these three cases of removing points, the reduced charge matrix is the same as given by eq. (\ref{Q-new-P3}) and we get the same diagrams as shown in figure \ref{Phase-I-C5-INTO-P3}.

\section{Embedding of Fano $\Ba$ inside Fano $\Cb$}
\label{sec4}
\begin{figure}[tbp]
	\centering
		\includegraphics[width=0.90\textwidth]{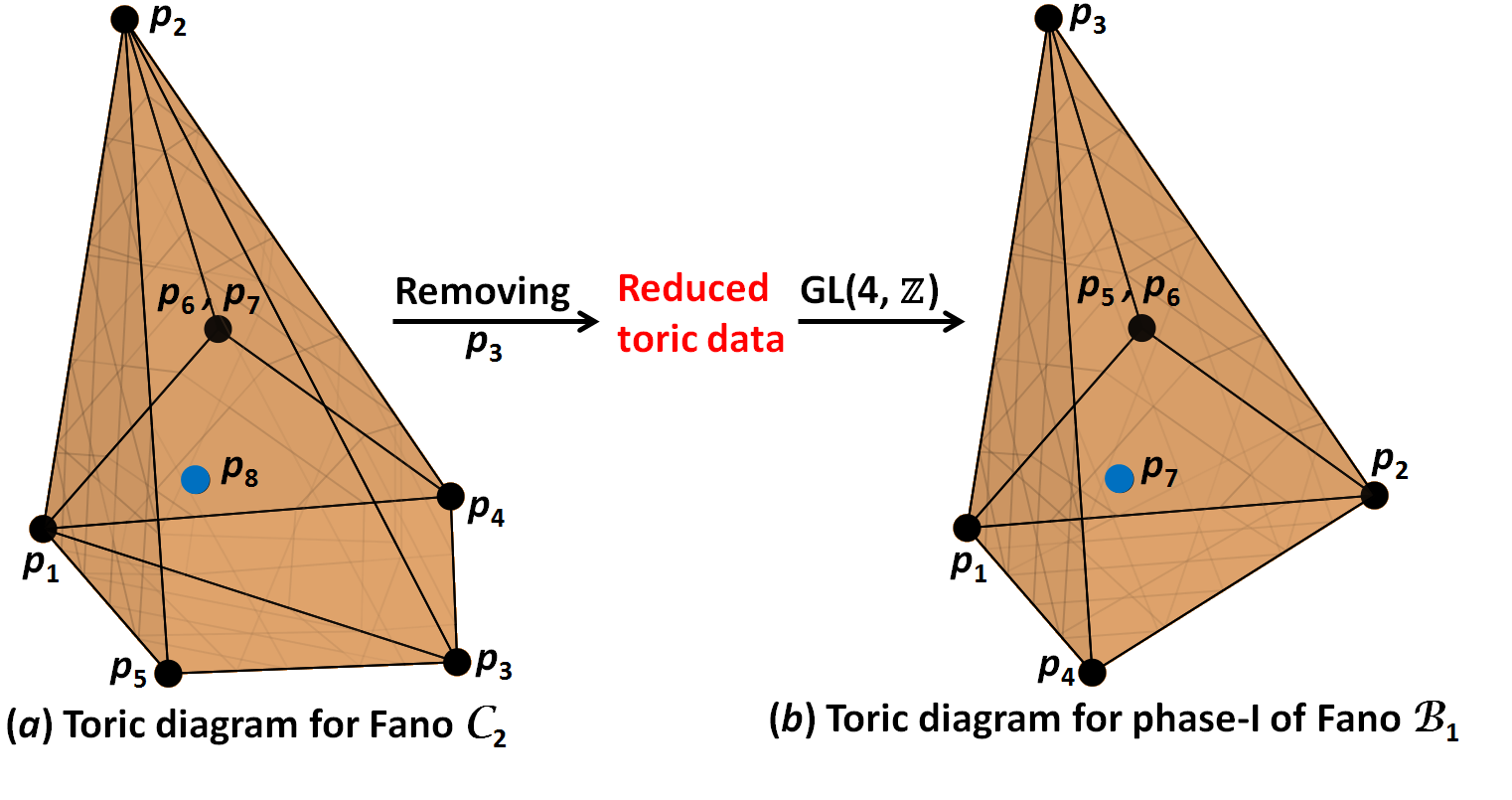}
	\caption{Toric diagram for Fano $\Cb$ is shown in figure ($a$). Removing $p_3$ will give a reduced toric data which is equivalent to toric diagram for phase-I of Fano $\Ba$ given in figure ($b$).}
	\label{ToricC2toB1}
\end{figure}

In this section, we will discuss about the embeddings of Fano $\Ba$ inside Fano $\Cb$ and also the partial resolution of Fano $\Cb$. 
The quiver gauge theory corresponding to Fano $\Cb$ is discussed in \cite{Davey:2011mz}. The information about the quiver gauge theory can be encoded in the charge matrix (Q) which is given by \cite{Davey:2011mz}:
\begin{equation}
Q_{\Cb} =\left(
\begin{array}{c}
 Q_F \\ \hline
 Q_D
\end{array}
\right)=\left(
\begin{array}{cccccccc}
 1 & 1 & 1 & 0 & -1 & -1 & -1 & 0 \\
 0 & 0 & 1 & -1 & -1 & 0 & 0 & 1 \\ \hline
 0 & 0 & 0 & 0 & 1 & 1 & 0 & -2 \\
 0 & 0 & 0 & 0 & 0 & 1 & -1 & 0
\end{array}
\right) ~.
\label{Q-C2}
\end{equation}
The toric data for Fano $\Cb$ corresponding to this quiver gauge theory is given as \cite{Davey:2011mz}:
\begin{equation}
{\cal{G}}_{\Cb} =\left(
\begin{array}{cccccccc}
 p_1 & p_2 & p_3 & p_4 & p_5 & p_6 & p_7 & p_8 \\ \hline
 1 & 1 & 1 & 1 & 1 & 1 & 1 & 1 \\
 1 & -1 & 0 & 0 & 0 & 0 & 0 & 0 \\
 0 & 1 & -1 & -1 & 0 & 0 & 0 & 0 \\
 1 & 0 & 0 & 1 & -1 & 1 & 1 & 0
\end{array}
\right)~.
\label{G-C2}
\end{equation}
The toric diagram encoded by (\ref{G-C2}) is shown in figure \ref{ToricC2toB1}($a$).
There are two possible quiver Chern-Simons theories corresponding to complex cone over Fano $\Ba$ which were obtained in \cite{Dwivedi:2011zm} and \cite{Phukon:2011hp}. We will call them as phase-I and phase-II of Fano $\Ba$ respectively. 
The charge matrix for phase-I of Fano $\Ba$ and the corresponding toric data are given as \cite{Dwivedi:2011zm}:
\begin{eqnarray}
Q^{phase-I}_{\Ba} & = & \left(
\begin{array}{c}
 Q_F \\ \hline
 Q_D
\end{array}
\right)=\left(
\begin{array}{ccccccc}
 1 & 1 & 1 & -2 & -2 & -2 & 3 \\
 0 & 0 & 0 & 2 & 1 & 1 & -4 \\ \hline
 0 & 0 & 0 & 1 & 0 & 1 & -2
\end{array}
\right) ; \nonumber \\ 
{\cal{G}}^{phase-I}_{\Ba} & = & \left(
\begin{array}{ccccccc}
p_1 & p_2 & p_3 & p_4 & p_5 & p_6 & p_7 \\ \hline
 1 & 1 & 1 & 1 & 1 & 1 & 1 \\
 1 & -1 & 0 & 0 & 0 & 0 & 0 \\
 0 & 1 & -1 & 0 & 0 & 0 & 0 \\
 0 & 0 & 2 & -1 & 1 & 1 & 0
\end{array}
\right)~.
\label{Q-G-Phase-I-B1}
\end{eqnarray}
Similarly, the charge matrix for phase-II of Fano $\Ba$ and the corresponding toric data are given below \cite{Phukon:2011hp}:
\begin{eqnarray}
Q^{phase-II}_{\Ba}  & = & \left(
\begin{array}{c}
 Q_F \\ \hline
 Q_D
\end{array}
\right)  = 
\left(
\begin{array}{cccccccc}
 1 & 1 & 1 & 0 & -1 & 0 & -1 & -1  \\
 0 & 0 & 0 & 0 & 0 & 1 & 0 & -1 \\ \hline
 0 & 0 & 0 & 1 & 0 & -1 & 1 & -1 \\ 
 0 & 0 & 0 & 0 & 1 & 0 & -1 & 0 
\end{array}
\right); \nonumber \\ 
{\cal{G}}^{phase-II}_{\Ba} & = & \left(
\begin{array}{cccccccc}
p_1 & p_2 & p_3 & p_4 & p_5 & p_6 & p_7 & p_8 \\ \hline
 1 & 1 & 1 & 1 & 1 & 1 & 1 & 1  \\
 1 & -1 & 0 & 0 & 0 & 0 & 0 & 0  \\
 0 & 1 & -1 & 0 & 0 & 0 & 0 & 0 \\
 0 & 0 & 2 & -1 & 1 & 0 & 1 & 0
\end{array}
\right)~.
\label{Q-G-Phase-II-B1}
\end{eqnarray}
We can see that the toric data of both the phases are same except for the different multiplicities of the point ($0,0,0$). The toric diagram for phase-I of Fano $\Ba$ is given in figure \ref{ToricC2toB1}($b$) where the point ($0,0,0$) is indicated by $p_7$. For phase-II of  Fano $\Ba$, the point ($0,0,0$) will correspond to $p_7, p_8$. Since we have different number of columns in the two phases of Fano $\Ba$, we must study the embedding and partial resolution of both the phases separately. 

\subsection{Partial resolution of Fano $\Cb$ to phase-I of Fano $\Ba$}
If we remove the third column ($p_3$) from the toric data of Fano $\Cb$ (\ref{G-C2}), the new toric data is $GL(4,\BZ)$ related to the toric data of phase-I of Fano $\Ba$ (\ref{Q-G-Phase-I-B1}) as given below:
\begin{equation}
{\cal{G}}_{\Ba}^{phase-I} =
\left(
\begin{array}{cccc}
 1 & 0 & 0 & 0 \\
 0 & 1 & 0 & 0 \\
 0 & 0 & 1 & 0 \\
 0 & -1 & -1 & 1
\end{array}
\right).\left(
\begin{array}{ccccccc}
 p_1 & p_2 & p_4 & p_5 & p_6 & p_7 & p_8 \\ \hline
 1 & 1 & 1 & 1 & 1 & 1 & 1 \\
 1 & -1 & 0 & 0 & 0 & 0 & 0 \\
 0 & 1 & -1 & 0 & 0 & 0 & 0 \\
 1 & 0 & 1 & -1 & 1 & 1 & 0
\end{array}
\right)~.
\end{equation}
Thus we find that the toric data for phase-I of Fano $\Ba$ is embedded inside that of Fano $\Cb$. Next we find the reduced charge matrix obtained as a result of removal of the point $p_3$ from the toric data of Fano $\Cb$. The procedure is same as given earlier. A row ($r$) of the reduced charge matrix (which is to be obtained) for Fano $\Ba$ can be written as linear combination of the rows $R_1, R_2, R_3, R_4$ of charge matrix for Fano $\Cb$ given in eq. (\ref{Q-C2}):
\begin{eqnarray}
r & = & a_1R_1 + a_2R_2 + a_3R_3 + a_4R_4 \nonumber \\ 
& = & \left( a_1, a_1, a_1+a_2, -a_2,  -a_1-a_2+a_3, \right. \nonumber \\
& & \left. -a_1+a_3+a_4, -a_1-a_4, a_2-2 a_3  \right) ~.
\label{lin-comb-C2}
\end{eqnarray}
Now, since we removed $p_3$ from the toric data, we must set the third column to 0 in the eq. (\ref{lin-comb-C2}), which means setting $a_1+a_2 =0$. This gives $a_2 = -a_1$. Substituting it back in the linear combination, we get,
\begin{eqnarray}
r & = & a_1R_1 - a_1R_2 + a_3R_3 + a_4R_4 \nonumber \\ 
& = & a_1\left(R_1 - R_2\right) + a_3R_3 + a_4R_4  
\end{eqnarray}
Thus we find that the reduced charge matrix (say $Q$) is spanned by three rows: $(R_1 - R_2)$ which will form a $Q_F$ row and the two rows $R_3$ and $R_4$, both of which will be rows of $Q_D$ charge matrix. Thus the final reduced charge matrix can be written as: 
\begin{equation}
Q =\left(
\begin{array}{c}
 Q_F \\ \hline
 Q_D
\end{array}
\right)=\left(
\begin{array}{ccccccc}
 1 & 1 & 1 & 0 & -1 & -1 & -1 \\ \hline
 0 & 0 & 0 & 1 & 1 & 0 & -2 \\
 0 & 0 & 0 & 0 & 1 & -1 & 0
\end{array}
\right) ~,
\label{Q-reduced-B1}
\end{equation}
where we have deleted the entire third column in the expression of $Q$ (which was already set to 0).
This reduced charge matrix $Q$ is different from the charge matrices (\ref{Q-G-Phase-I-B1}) and (\ref{Q-G-Phase-II-B1}) of the known quiver gauge theories for Fano $\Ba$. So we must check if this charge matrix ($Q$) encodes a quiver diagram. For this, we must apply the inverse algorithm starting from the charge matrix $Q$ in eq. (\ref{Q-reduced-B1}). The perfect matching matrix from (\ref{Q-reduced-B1}) comes out to be,
\begin{equation}
P = \left(
\begin{array}{c|ccccccc}
& p_1 & p_2 & p_3 & p_4 & p_5 & p_6 & p_7 \\ \hline
X_1 & 0 & 0 & 0 & 1 & 0 & 0 & 0 \\
X_2 & 1 & 0 & 0 & 0 & 1 & 0 & 0 \\
X_3 & 0 & 1 & 0 & 0 & 1 & 0 & 0 \\
X_4 & 0 & 0 & 1 & 0 & 1 & 0 & 0 \\
X_5 & 1 & 0 & 0 & 0 & 0 & 1 & 0 \\
X_6 & 0 & 1 & 0 & 0 & 0 & 1 & 0 \\
X_7 & 0 & 0 & 1 & 0 & 0 & 1 & 0 \\
X_8 & 1 & 0 & 0 & 0 & 0 & 0 & 1 \\
X_9 & 0 & 1 & 0 & 0 & 0 & 0 & 1 \\
X_{10} & 0 & 0 & 1 & 0 & 0 & 0 & 1 \\
\end{array}
\right)
\end{equation}
We find that this $P$ matrix does not give any sensible quiver diagram.  

In case of phase-II of Fano $\Ba$, we see that the toric data has 8 columns as given in (\ref{Q-G-Phase-II-B1}), where the vertices ($0,0,0$) and ($0,0,1$) both having multiplicities 2. Moreover, the toric data for Fano $\Cb$ also has 8 columns given by (\ref{G-C2}) where only vertex ($0,0,1$) has multiplicity 2. We could not find any $GL(4,\BZ)$ matrix, by which the two toric datas can be related. Hence,  we could not perform the partial resolution in this case. 

\section{Embedding of Fano $\Bd$ inside other Fano threefolds }
\label{sec5}
In this section, we will discuss about the possible embeddings of Fano $\Bd$ inside the remaining 17 Fano threefolds. We tried all the possibilities of removing points from the toric diagrams of these 17 Fano threefolds to get the toric diagram of Fano $\Bd$. We find that Fano $\Bd$ can be embedded inside Fano $\Db$ and Fano $\Cd$. Note that the embedding of Fano $\Bd$ inside Fano $\Db$ and Fano $\Cd$ has already been seen in \cite{Phukon:2011hp}. In the following subsections, we will discuss these embeddings and then use the method of partial resolution to obtain the possible reduced charge matrices for Fano $\Bd$. We will then apply inverse algorithm on these reduced charge matrices to find any quiver diagram.
 
\subsection{Embedding of Fano $\Bd$ inside Fano $\Db$}
\begin{figure}[tbp]
	\centering
		\includegraphics[width=1.05\textwidth]{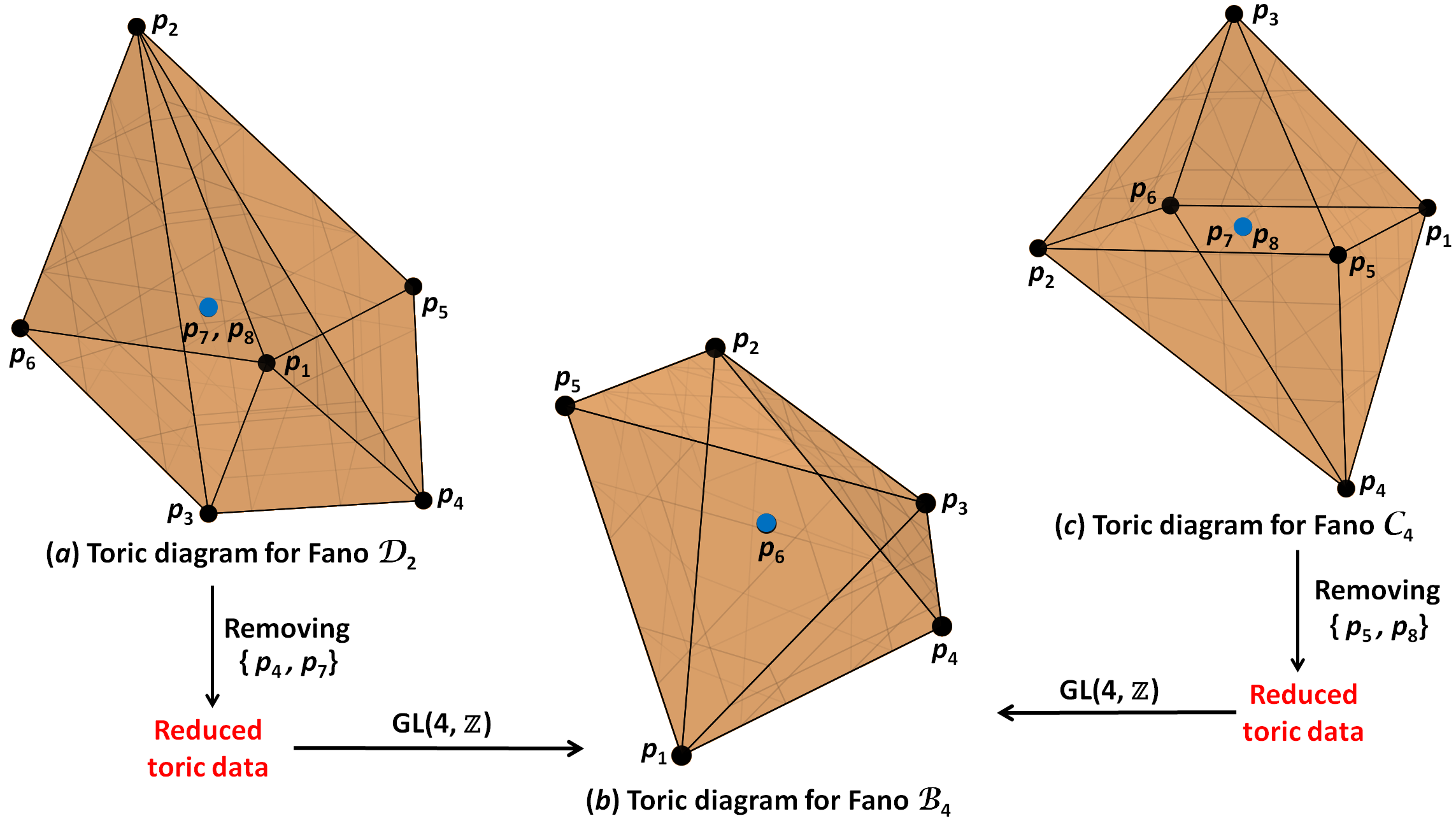}
	\caption{Toric diagram for Fano $\Db$ is shown in figure ($a$). Removing $\{p_4,p_7\}$ or $\{p_4,p_8\}$ will give the toric data for Fano $\Bd$ given in figure ($b$). Figure ($c$) is the toric diagram for Fano $\Cd$ and removing $\{p_5,p_7\}$ or $\{p_5,p_8\}$ will give Fano $\Bd$ toric diagram.}
	\label{ToricD2C4toB4}
\end{figure}

\begin{figure}
	\centering
		\includegraphics[width=0.50\textwidth]{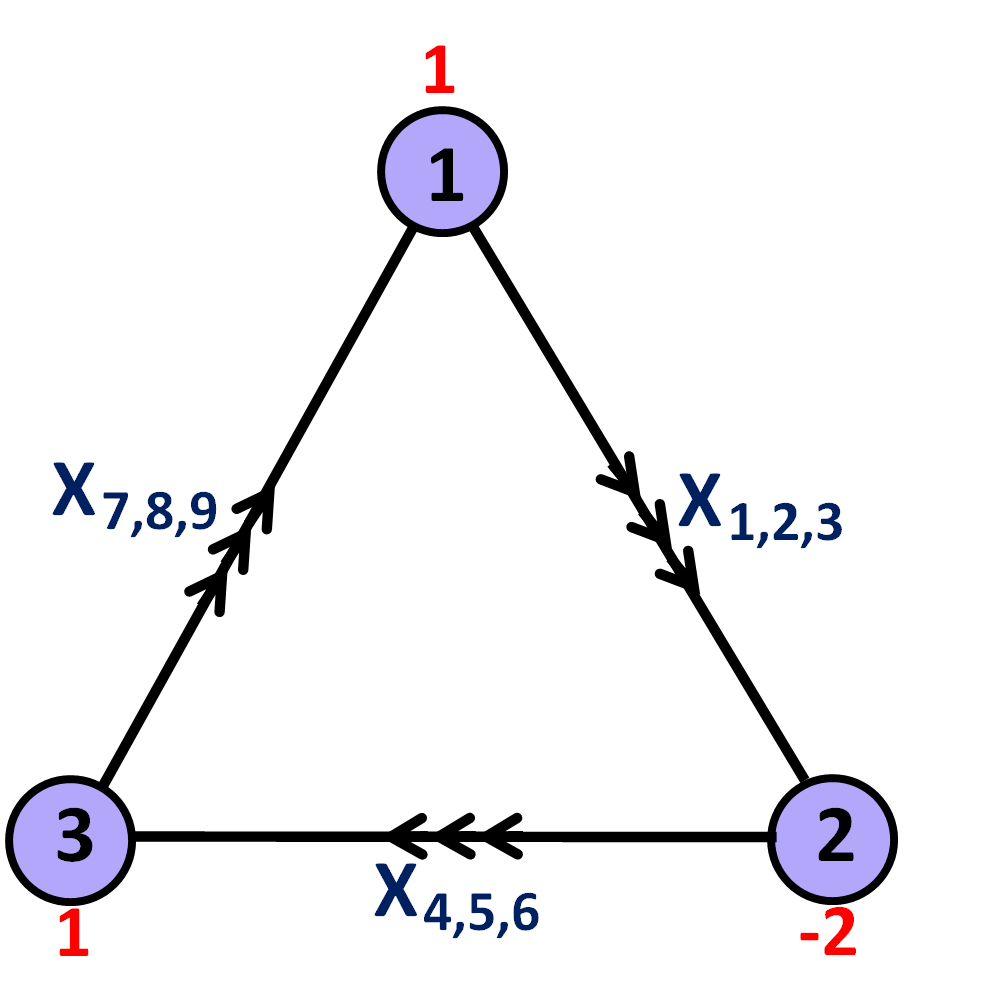}
	\caption{Quiver diagram and Chern-Simons levels obtained for Fano $\Bd$ from reduced charge matrix (\ref{Q-reduced-B4}), as a result of partial resolution of Fano $\Db$. The quiver diagram matches with the known quiver for Fano $\Bd$ \cite{Davey:2011mz}.}
	\label{D2-INTO-B4}
\end{figure}

The toric data of Fano $\Bd$ can be embedded inside the toric data of Fano $\Db$ \cite{Phukon:2011hp}. The quiver gauge theory corresponding to Fano $\Db$ is discussed in \cite{Davey:2011mz}. The information about the quiver gauge theory can be encoded in the charge matrix (Q) which is given by \cite{Davey:2011mz}:
\begin{equation}
Q_{\Db} =\left(
\begin{array}{c}
 Q_F \\ \hline
 Q_D
\end{array}
\right)=\left(
\begin{array}{cccccccc}
 1 & 1 & 0 & 1 & -1 & 0 & -1 & -1 \\
 0 & 0 & 1 & -1 & 0 & -1 & 1 & 0 \\ \hline
 0 & 0 & 0 & 0 & 1 & 1 & 0 & -2 \\
 0 & 0 & 0 & 0 & 0 & 0 & 1 & -1
\end{array}
\right) ~.
\label{Q-D2-first}
\end{equation}
The toric data for Fano $\Db$ corresponding to this quiver gauge theory is given as \cite{Davey:2011mz}:
\begin{equation}
{\cal{G}}_{\Db} =\left(
\begin{array}{cccccccc}
 p_1 & p_2 & p_3 & p_4 & p_5 & p_6 & p_7 & p_8 \\ \hline
 1 & 1 & 1 & 1 & 1 & 1 & 1 & 1 \\
 1 & -1 & 0 & 0 & 0 & 0 & 0 & 0 \\
 0 & 1 & -1 & -1 & 0 & 0 & 0 & 0 \\
 0 & 0 & 0 & 1 & 1 & -1 & 0 & 0
\end{array}
\right)~.
\label{G-D2-first}
\end{equation}
The toric diagram is shown in figure \ref{ToricD2C4toB4}($a$).
This toric diagram embeds the toric diagram of Fano $\Bd$ shown in figure \ref{ToricD2C4toB4}($b$), the vertices of which can be encoded in the following toric data \cite{Dwivedi:2011zm}:
\begin{equation}
{\cal{G}}_{\Bd} = \left(
\begin{array}{cccccc}
p_1 & p_2 & p_3 & p_4 & p_5 & p_6 \\ \hline
 1 & 1 & 1 & 1 & 1 & 1 \\
 1 & -1 & 0 & 0 & 0 & 0 \\
 0 & 1 & -1 & 0 & 0 & 0 \\
 0 & 0 & 0 & 1 & -1 & 0
\end{array}
\right)~.
\label{G-B4}
\end{equation}
The quiver gauge theory for Fano $\Bd$ was given in \cite{Davey:2011mz} and the corresponding charge matrix is given as:
\begin{equation}
Q_{\Bd} =\left(
\begin{array}{c}
 Q_F \\ \hline
 Q_D
\end{array}
\right)=\left(
\begin{array}{cccccc}
 1 & 1 & 1 & -1 & -1 & -1 \\ \hline
 0 & 0 & 0 & 1 & 1 & -2
\end{array}
\right) ~.
\label{Q-B4}
\end{equation}
We see that the toric data of Fano $\Bd$ given in (\ref{G-B4}) is related by a $GL(4,\BZ)$ transformation to the toric data of Fano $\Db$ (\ref{G-D2-first}) if we remove the set of points $\{p_4, p_7 \}$ or $\{p_4, p_8 \}$ from (\ref{G-D2-first}) as given below:
\begin{eqnarray}
{\cal{G}}_{\Bd} & = &
\left(
\begin{array}{cccc}
 1 & 0 & 0 & 0 \\
 0 & 1 & 0 & 0 \\
 0 & 0 & 1 & 0 \\
 0 & 0 & 0 & 1
\end{array}
\right).\left(
\begin{array}{cccccc}
 p_1 & p_2 & p_3 & p_5 & p_6 & p_8 \\ \hline
 1 & 1 & 1 & 1 & 1 & 1 \\
 1 & -1 & 0 & 0 & 0 & 0 \\
 0 & 1 & -1 & 0 & 0 & 0 \\
 0 & 0 & 0 & 1 & -1 & 0
\end{array}
\right) \nonumber \\
{\cal{G}}_{\Bd} & = &
\left(
\begin{array}{cccc}
 1 & 0 & 0 & 0 \\
 0 & 1 & 0 & 0 \\
 0 & 0 & 1 & 0 \\
 0 & 0 & 0 & 1
\end{array}
\right).\left(
\begin{array}{cccccc}
 p_1 & p_2 & p_3 & p_5 & p_6 & p_7 \\ \hline
 1 & 1 & 1 & 1 & 1 & 1 \\
 1 & -1 & 0 & 0 & 0 & 0 \\
 0 & 1 & -1 & 0 & 0 & 0 \\
 0 & 0 & 0 & 1 & -1 & 0
\end{array}
\right) 
\end{eqnarray}
We start with the choice of removing $\{p_4, p_7 \}$ points from the toric diagram of Fano $\Db$ and find the reduced charge matrix obtained as a result of removal of these points. 
A row ($r$) of the reduced charge matrix (which is to be obtained) can be written as linear combination of the rows $R_1, R_2, R_3, R_4$ of charge matrix for Fano $\Db$ given in eq. (\ref{Q-D2-first}):
\begin{eqnarray}
r & = & a_1R_1 + a_2R_2 + a_3R_3 + a_4R_4 \nonumber \\ 
& = & \left( a_1, a_1, a_2, a_1-a_2,  -a_1+a_3, -a_2+a_3, \right. \nonumber \\
&& \left. -a_1+a_2+a_4, -a_1-2a_3-a_4  \right)
\label{lin-comb-D2}
\end{eqnarray}
Removing $\{p_4, p_7 \}$ columns, we must set the corresponding columns in (\ref{lin-comb-D2}) to 0, which means setting $a_1-a_2=0$ and $-a_1+a_2+a_4 = 0$. This gives $a_2=a_1$ and $a_4=0$. Substituting it back in the linear combination, we get,
\begin{eqnarray}
r & = & a_1R_1 + a_1R_2 + a_3R_3 \nonumber \\ 
& = & a_1\left(R_1 + R_2\right) + a_3R_3
\end{eqnarray}
Thus we find that the reduced charge matrix (say $Q$) is spanned by two rows: $(R_1 + R_2)$ which will form a $Q_F$ row and $R_3$, which will form a $Q_D$ row. Thus the final reduced charge matrix can be written as (removing fourth and seventh columns):  
\begin{equation}
Q =\left(
\begin{array}{c}
 Q_F \\ \hline
 Q_D
\end{array}
\right)=\left(
\begin{array}{cccccc}
 1 & 1 & 1 & -1 & -1 & -1 \\ \hline
 0 & 0 & 0 & 1 & 1 & -2
\end{array}
\right) ~.
\label{Q-reduced-B4}
\end{equation}
The $T$ and $K$ matrices are given as,
\begin{equation}
T = \left(
\begin{array}{cccccc}
 1 & 0 & 0 & 0 & 0 & 1 \\
 1 & 0 & 0 & 0 & 1 & 0 \\
 1 & 0 & 0 & 1 & 0 & 0 \\
 -1 & 0 & 1 & 0 & 0 & 0 \\
 -1 & 1 & 0 & 0 & 0 & 0 \\
\end{array}
\right)~; K = \left(
\begin{array}{ccccc}
 0 & 0 & 1 & 0 & 0 \\
 0 & 0 & 1 & 0 & 1 \\
 0 & 0 & 1 & 1 & 0 \\
 0 & 1 & 0 & 0 & 0 \\
 0 & 1 & 0 & 0 & 1 \\
 0 & 1 & 0 & 1 & 0 \\
 1 & 0 & 0 & 0 & 0 \\
 1 & 0 & 0 & 0 & 1 \\
 1 & 0 & 0 & 1 & 0 \\
\end{array}
\right)~.
\end{equation}
The perfect matching matrix is given as,
\begin{equation}
P = \left(
\begin{array}{c|cccccc}
& p_1 & p_2 & p_3 & p_4 & p_5 & p_6 \\ \hline
X_1 & 1 & 0 & 0 & 1 & 0 & 0 \\
X_2 & 0 & 1 & 0 & 1 & 0 & 0 \\
X_3 & 0 & 0 & 1 & 1 & 0 & 0 \\
X_4 & 1 & 0 & 0 & 0 & 1 & 0 \\
X_5 & 0 & 1 & 0 & 0 & 1 & 0 \\
X_6 & 0 & 0 & 1 & 0 & 1 & 0 \\
X_7 & 1 & 0 & 0 & 0 & 0 & 1 \\
X_8 & 0 & 1 & 0 & 0 & 0 & 1 \\
X_9 & 0 & 0 & 1 & 0 & 0 & 1 \\
\end{array}
\right)~.
\end{equation}
From this $P$ matrix we can obtain the quiver diagram and the Chern-Simons levels as shown in figure \ref{D2-INTO-B4}.
The Chern-Simons levels $(k_1, k_2, k_3) = (1, -2, 1)$ have been written in red across the nodes of the quiver \ref{D2-INTO-B4}. The superpotential of the theory can be constructed from $K$ and comes out to be:
\begin{equation}
W_{\Bd} = X_1 X_5 X_9 - X_1 X_6 X_8 + X_2 X_6 X_7 - X_2 X_4 X_9 + X_3 X_4 X_8 - X_3 X_5 X_7 ~.
\end{equation}
We see that this is the same quiver gauge theory for Fano $\Bd$ as given in \cite{Davey:2011mz}. In \cite{Davey:2011mz}, a quiver gauge theory (obtained from the brane tiling approach \cite{Franco:2005rj, Hanany:2005ve,Hanany:2008fj,Davey:2009bp}) was selected and forward algorithm was used to verify that this was the correct theory for Fano $\Bd$. Here, we used the partial resolution approach and resolved Fano $\Db$, to get a quiver Chern-Simons theory for Fano $\Bd$, which happens to be the same as given in \cite{Davey:2011mz}. 

We also tried removing the points $\{p_4, p_8 \}$ from the toric diagram of Fano $\Db$. The reduced charge matrix comes out to be:
\begin{equation}
Q =  Q_D
=\left(
\begin{array}{cccccc}
 1 & 1 & 1 & -1 & -1 & -1 \\
 0 & 0 & 0 & 1 & 1 & -2
\end{array}
\right) ~.
\label{Q-reduced-B4-second}
\end{equation}
The perfect matching matrix is,
\begin{equation}
P = \left(
\begin{array}{c|cccccc}
& p_1 & p_2 & p_3 & p_4 & p_5 & p_6 \\ \hline
X_1 & 1 & 0 & 0 & 0 & 0 & 0 \\
X_2 & 0 & 1 & 0 & 0 & 0 & 0 \\
X_3 & 0 & 0 & 1 & 0 & 0 & 0 \\
X_4 & 0 & 0 & 0 & 1 & 0 & 0 \\
X_5 & 0 & 0 & 0 & 0 & 1 & 0 \\
X_6 & 0 & 0 & 0 & 0 & 0 & 1 \\
\end{array}
\right)
\end{equation}
This $P$ matrix does not give any quiver diagram.

\subsection{Embedding of Fano $\Bd$ inside Fano $\Cd$}
The quiver gauge theory corresponding to Fano $\Cd$ is discussed in \cite{Davey:2011mz}. The information about the quiver gauge theory can be encoded in the charge matrix (Q) given by \cite{Davey:2011mz}:
\begin{equation}
Q_{\Cd} =\left(
\begin{array}{c}
 Q_F \\ \hline
 Q_D
\end{array}
\right)=\left(
\begin{array}{cccccccc}
 1 & 1 & -1 & -1 & 0 & -1 & 1 & 0 \\
 0 & 0 & 0 & 0 & 1 & 1 & -1 & -1 \\ \hline
 1 & 1 & 0 & 0 & 0 & 0 & 0 & -2 \\
 0 & 0 & 0 & 0 & 0 & 0 & 1 & -1
\end{array}
\right) ~.
\label{Q-C4-sid}
\end{equation}
The toric data for Fano $\Cd$ corresponding to this quiver gauge theory is given as \cite{Davey:2011mz}:
\begin{equation}
{\cal{G}}_{\Cd} =\left(
\begin{array}{cccccccc}
 p_1 & p_2 & p_3 & p_4 & p_5 & p_6 & p_7 & p_8 \\ \hline
 1 & 1 & 1 & 1 & 1 & 1 & 1 & 1 \\
 1 & -1 & 0 & 0 & 0 & 0 & 0 & 0 \\
 0 & 0 & 1 & -1 & 0 & 0 & 0 & 0 \\
 0 & 0 & 0 & 1 & 1 & -1 & 0 & 0
\end{array}
\right)~.
\label{G-C4-sid}
\end{equation}
The toric diagram is shown in figure \ref{ToricD2C4toB4}($c$). This toric data embeds the toric diagram of Fano $\Bd$ \cite{Phukon:2011hp}, if we remove the columns $\left\{p_5, p_7\right\}$ or $\left\{p_5, p_8\right\}$ from the toric diagram of Fano $\Cd$ given by (\ref{G-C4-sid}) as shown in figure \ref{ToricD2C4toB4}. Using the partial resolution method, as discussed earlier, we write a row of charge matrix of Fano $\Bd$ (to be obtained) as linear combination of rows of charge matrix (\ref{Q-C4-sid}) of Fano $\Cd$ and delete the $\left\{p_5, p_8\right\}$ columns. Finding the linear combination, we get the following reduced charge matrix for Fano $\Bd$:
\begin{equation}
Q =\left(
\begin{array}{c}
 Q_F \\ \hline
 Q_D
\end{array}
\right)=\left(
\begin{array}{cccccc}
 1 & 1 & -1 & -1 & -1 & 1 \\ \hline
 1 & 1 & 0 & 0 & 0 & -2
\end{array}
\right) ~.
\label{Q-reduced-B4-third}
\end{equation}
This charge matrix is different from the charge matrices of Fano $\Bd$ given in (\ref{Q-reduced-B4}) and (\ref{Q-reduced-B4-second}) and gives a $P$ matrix as,
\begin{equation}
P = \left(
\begin{array}{c|cccccc}
& p_1 & p_2 & p_3 & p_4 & p_5 & p_6 \\ \hline
X_1 & 1 & 0 & 1 & 0 & 0 & 0 \\
X_2 & 0 & 1 & 1 & 0 & 0 & 0 \\
X_3 & 1 & 0 & 0 & 1 & 0 & 0 \\
X_4 & 0 & 1 & 0 & 1 & 0 & 0 \\
X_5 & 1 & 0 & 0 & 0 & 1 & 0 \\
X_6 & 0 & 1 & 0 & 0 & 1 & 0 \\
X_7 & 0 & 0 & 1 & 0 & 0 & 1 \\
X_8 & 0 & 0 & 0 & 1 & 0 & 1 \\
X_9 & 0 & 0 & 0 & 0 & 1 & 1 \\
\end{array}
\right)
\end{equation}
 However we checked that this does not give any quiver diagram.

\section{Embedding of Fano $\Ca$ inside Fano $\Eb$ }
\label{sec6}
\begin{figure}[tbp]
	\centering
		\includegraphics[width=1.05\textwidth]{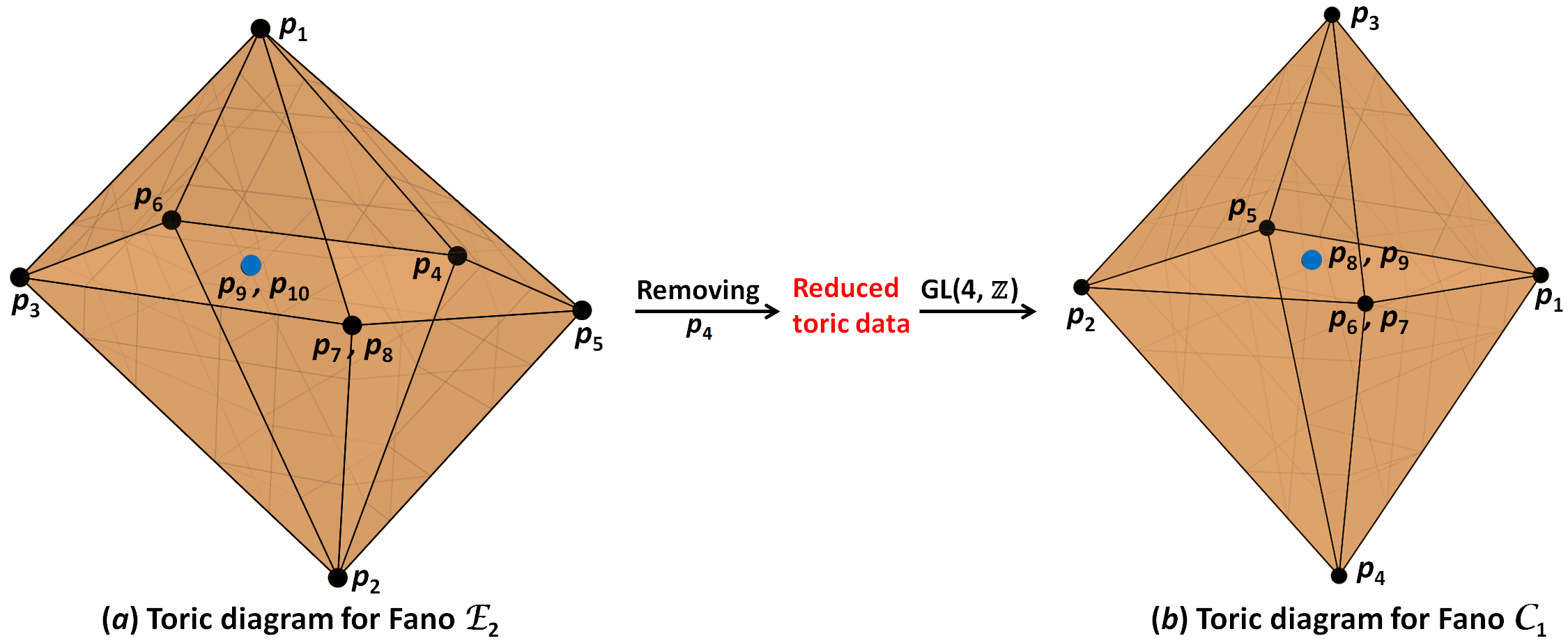}
	\caption{Toric diagram for Fano $\Eb$ is shown in figure ($a$). Removing $p_4$ will give the toric data for Fano $\Ca$ given in figure ($b$).}
	\label{ToricE2toC1}
\end{figure}

In this section, we will discuss about the embedding of Fano $\Ca$ inside Fano $\Eb$. The quiver gauge theory corresponding to Fano $\Eb$ is discussed in \cite{Davey:2011mz}. The information about the quiver gauge theory can be encoded in the charge matrix (Q) which is given by \cite{Davey:2011mz}:
\begin{equation}
Q_{\Eb} =\left(
\begin{array}{c}
 Q_F \\ \hline
 Q_D
\end{array}
\right)=\left(
\begin{array}{cccccccccc}
 1 & 1 & 0 & 0 & 0 & 0 & 0 & -1 & -1 & 0 \\
 0 & 0 & 1 & 1 & 0 & -1 & -1 & 0 & 0 & 0 \\
 0 & 0 & 0 & 1 & -1 & -1 & 0 & 0 & 0 & 1 \\ \hline
 0 & 0 & 0 & 0 & 0 & 1 & 1 & 0 & -2 & 0 \\
 0 & 0 & 0 & 0 & 0 & 0 & 1 & -1 & 0 & 0 \\
 0 & 0 & 0 & 0 & 0 & 0 & 0 & 0 & 1 & -1
\end{array}
\right) ~.
\label{Q-E2}
\end{equation}
The toric data for Fano $\Eb$ corresponding to this quiver gauge theory is given as \cite{Davey:2011mz}:
\begin{equation}
{\cal{G}}_{\Eb} =\left(
\begin{array}{cccccccccc}
 p_1 & p_2 & p_3 & p_4 & p_5 & p_6 & p_7 & p_8 & p_9 & p_{10} \\ \hline
 1 & 1 & 1 & 1 & 1 & 1 & 1 & 1 & 1 & 1 \\
 1 & -1 & 0 & 0 & 0 & 0 & 0 & 0 & 0 & 0 \\
 0 & 0 & 1 & -1 & -1 & 0 & 0 & 0 & 0 & 0 \\
 0 & 1 & 0 & 0 & 1 & -1 & 1 & 1 & 0 & 0
\end{array}
\right)~.
\label{G-E2}
\end{equation}
The toric diagram is  shown in figure \ref{ToricE2toC1}($a$).
We find that the toric data of Fano $\Ca$ can only be embedded inside the toric data of Fano $\Eb$ (\ref{G-E2}). The vertices of the toric diagram of Fano $\Ca$ given in figure \ref{ToricE2toC1}($b$) can be encoded in terms of toric data given as \cite{Davey:2011mz}:
\begin{equation}
{\cal{G}}_{\Ca} = \left(
\begin{array}{ccccccccc}
p_1 & p_2 & p_3 & p_4 & p_5 & p_6 & p_7 & p_8 & p_9 \\ \hline
 1 & 1 & 1 & 1 & 1 & 1 & 1 & 1 & 1 \\
 1 & -1 & 0 & 0 & 0 & 0 & 0 & 0 & 0 \\
 0 & 0 & 1 & -1 & 0 & 0 & 0 & 0 & 0 \\
 0 & 1 & 0 & 1 & -1 & 1 & 1 & 0 & 0
\end{array}
\right)~.
\label{G-C1}
\end{equation}
The quiver gauge theory for Fano $\Ca$ was given in \cite{Davey:2011mz} and the corresponding charge matrix is given as:
\begin{equation}
Q_{\Ca} =
\left(
\begin{array}{c}
 Q_F \\ \hline
 Q_D
\end{array}
\right)=
\left(
\begin{array}{ccccccccc}
 1 & 1 & 0 & 0 & 0 & 0 & -1 & -1 & 0 \\
 0 & 0 & 1 & 1 & 0 & 0 & -1 & 0 & -1 \\
 0 & 0 & 0 & 0 & 1 & 1 & 0 & -1 & -1 \\ \hline
 0 & 0 & 0 & 0 & 0 & 1 & -1 & 0 & 0 \\
 0 & 0 & 0 & 0 & 0 & 0 & 0 & 1 & -1
\end{array}
\right) ~.
\label{Q-C1}
\end{equation}
We find that the toric data of Fano $\Ca$ given in (\ref{G-C1}) is related by a $GL(4,\BZ)$ transformation to the toric data of Fano $\Eb$ (\ref{G-E2}) if we remove $p_4$ as given below:
\begin{eqnarray}
{\cal{G}}_{\Ca}  = 
\left(
\begin{array}{cccc}
 1 & 0 & 0 & 0 \\
 0 & 1 & 0 & 0 \\
 0 & 0 & 1 & 0 \\
 0 & 0 & 0 & 1
\end{array}
\right).\left(
\begin{array}{ccccccccc}
 p_1 & p_2 & p_3 & p_5 & p_6 & p_7 & p_8 & p_9 & p_{10} \\ \hline
 1 & 1 & 1 & 1 & 1 & 1 & 1 & 1 & 1 \\
 1 & -1 & 0 & 0 & 0 & 0 & 0 & 0 & 0 \\
 0 & 0 & 1 & -1 & 0 & 0 & 0 & 0 & 0 \\
 0 & 1 & 0 & 1 & -1 & 1 & 1 & 0 & 0
\end{array}
\right) ~.  
\end{eqnarray}
The reduced charge matrix obtained as a result of removal of point $p_4$ from the toric diagram is given as:
\begin{equation}
Q =\left(
\begin{array}{c}
 Q_F \\ \hline
 Q_D
\end{array}
\right)=\left(
\begin{array}{ccccccccc}
 1 & 1 & 0 & 0 & 0 & 0 & -1 & -1 & 0 \\
 0 & 0 & 1 & 1 & 0 & -1 & 0 & 0 & -1 \\ \hline
 0 & 0 & 0 & 0 & 1 & 1 & 0 & -2 & 0 \\
 0 & 0 & 0 & 0 & 0 & 1 & -1 & 0 & 0 \\
 0 & 0 & 0 & 0 & 0 & 0 & 0 & 1 & -1
\end{array}
\right) ~.
\label{Q-reduced-C1}
\end{equation}
This reduced charge matrix is different from the charge matrix (\ref{Q-C1}) of the known quiver gauge theory for Fano $\Ca$ and gives the perfect matching matrix as,
\begin{equation}
P = \left(
\begin{array}{c|ccccccccc}
& p_1 & p_2 & p_3 & p_4 & p_5 & p_6 & p_7 & p_8 & p_9 \\ \hline
X_1 & 0 & 0 & 0 & 0 & 1 & 0 & 0 & 0 & 0 \\
X_2 & 0 & 0 & 1 & 0 & 0 & 1 & 0 & 0 & 0 \\
X_3 & 0 & 0 & 0 & 1 & 0 & 1 & 0 & 0 & 0 \\
X_4 & 1 & 0 & 0 & 0 & 0 & 0 & 1 & 0 & 0 \\
X_5 & 0 & 1 & 0 & 0 & 0 & 0 & 1 & 0 & 0 \\
X_6 & 1 & 0 & 0 & 0 & 0 & 0 & 0 & 1 & 0 \\
X_7 & 0 & 1 & 0 & 0 & 0 & 0 & 0 & 1 & 0 \\
X_8 & 0 & 0 & 1 & 0 & 0 & 0 & 0 & 0 & 1 \\
X_9 & 0 & 0 & 0 & 1 & 0 & 0 & 0 & 0 & 1 \\
\end{array}
\right)~.
\end{equation}
However, we found that it is not possible to encode this into any possible quiver diagram.

\section{Embedding of Fano $\Cc$ inside other Fano threefolds }
\label{sec7}
\begin{figure}[tbp]
	\centering
		\includegraphics[width=1.05\textwidth]{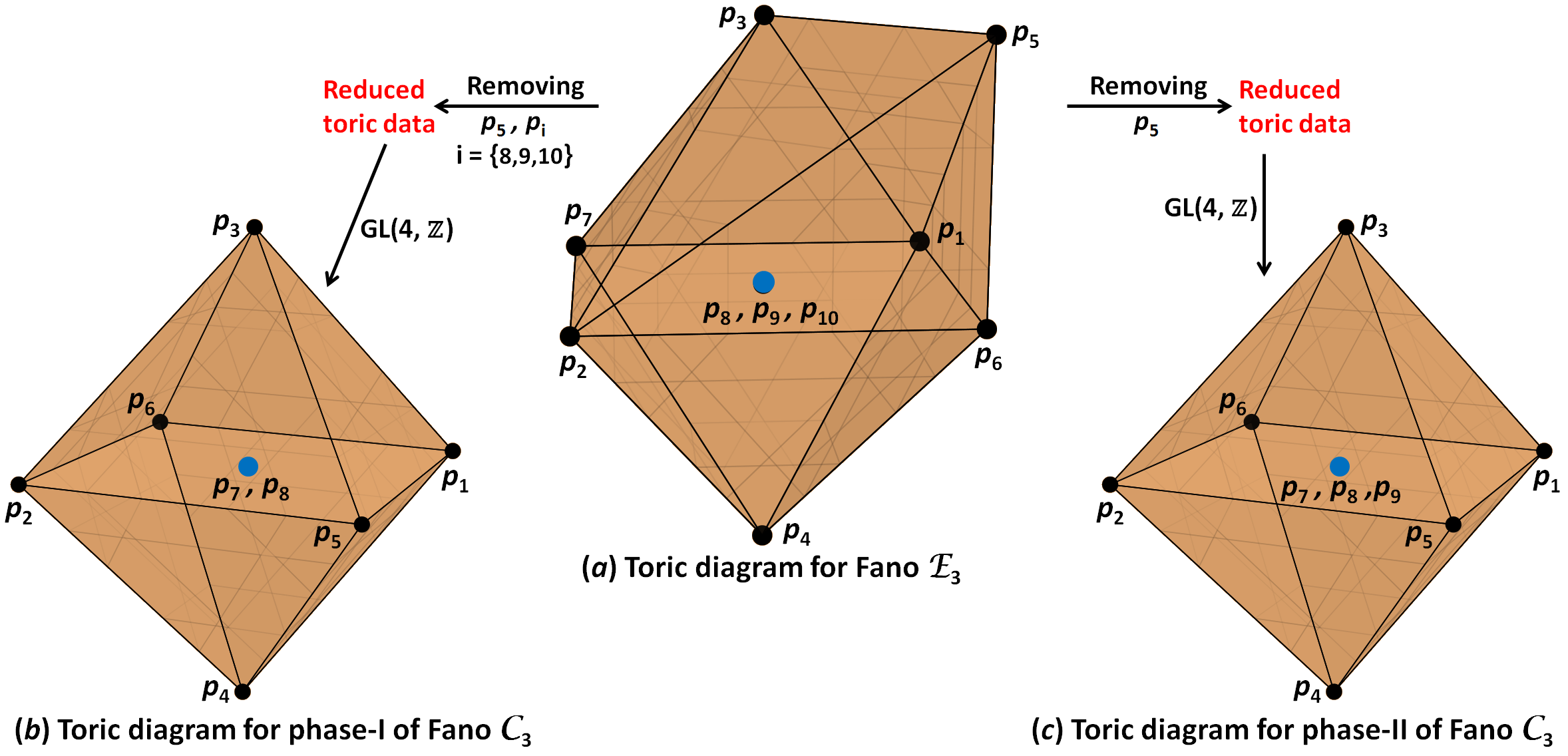}
	\caption{Toric diagram for Fano $\Ec$ is shown in figure ($a$). Removing $\{p_5,p_8\}$ or $\{p_5,p_9\}$ or $\{p_5,p_{10}\}$ will give the toric diagram for phase-I of Fano $\Cc$ given in figure ($b$). Simlarly removing only $p_5$ will give the toric diagram for phase-II of Fano $\Cc$ shown in figure ($c$).}
	\label{ToricE3tophaseIphaseIIC3}
\end{figure}

The Fano $\Cc$ theory is also known as the $Q^{1,1,1}/\BZ_2$ theory. There are two different quiver gauge theories known corresponding to the Calabi-Yau fourfold which is complex cone over Fano $\Cc$. These quiver theories are known as phase-I and phase-II of Fano $\Cc$ respectively and were obtained in \cite{Davey:2011mz}. The charge matrix for phase-I of Fano $\Cc$ and the corresponding toric data are given as:
\begin{eqnarray}
Q^{Phase-I}_{\Cc} & = & \left(
\begin{array}{c}
 Q_F \\ \hline
 Q_D
\end{array}
\right) =
\left(
\begin{array}{cccccccc}
 1 & 1 & 0 & 0 & -1 & -1 & 0 & 0 \\
 0 & 0 & 1 & 1 & 0 & 0 & -1 & -1 \\ \hline
 0 & 0 & 0 & 0 & 1 & 1 & -2 & 0 \\
 0 & 0 & 0 & 0 & 0 & 0 & 1 & -1
\end{array}
\right) ; \nonumber \\
{\cal{G}}^{Phase-I}_{\Cc} & = & \left(
\begin{array}{cccccccc}
p_1 & p_2 & p_3 & p_4 & p_5 & p_6 & p_7 & p_8 \\ \hline
 1 & 1 & 1 & 1 & 1 & 1 & 1 & 1 \\
 1 & -1 & 0 & 0 & 0 & 0 & 0 & 0 \\
 0 & 0 & 1 & -1 & 0 & 0 & 0 & 0 \\
 0 & 0 & 0 & 0 & 1 & -1 & 0 & 0
\end{array}
\right)~.
\end{eqnarray}
Similarly, the charge matrix for phase-II of Fano $\Cc$ and the corresponding toric data are given below:
\begin{eqnarray}
Q^{Phase-II}_{\Cc} & = & \left(
\begin{array}{c}
 Q_F \\ \hline
 Q_D
\end{array}
\right) =
\left(
\begin{array}{ccccccccc}
 1 & 1 & 0 & 0 & 0 & 0 & -1 & 0 & -1 \\
 0 & 0 & 1 & 1 & 0 & 0 & -1 & -1 & 0 \\
 0 & 0 & 0 & 0 & 1 & 1 & 0 & -1 & -1 \\ \hline
 0 & 0 & 0 & 0 & 1 & 1 & -2 & 0 & 0 \\
 0 & 0 & 0 & 0 & 0 & 0 & 1 & -1 & 0
\end{array}
\right); \nonumber \\ 
{\cal{G}}^{Phase-II}_{\Cc} & = & \left(
\begin{array}{ccccccccc}
p_1 & p_2 & p_3 & p_4 & p_5 & p_6 & p_7 & p_8 & p_9 \\ \hline
 1 & 1 & 1 & 1 & 1 & 1 & 1 & 1 & 1 \\
 1 & -1 & 0 & 0 & 0 & 0 & 0 & 0 & 0 \\
 0 & 0 & 1 & -1 & 0 & 0 & 0 & 0 & 0 \\
 0 & 0 & 0 & 0 & 1 & -1 & 0 & 0 & 0
\end{array}
\right)~.
\end{eqnarray}
The toric diagrams for both the phases I and II are shown in figure \ref{ToricE3tophaseIphaseIIC3}($b$) and \ref{ToricE3tophaseIphaseIIC3}($c$) respectively.
We can see that the toric data of both the phases are same except for the different multiplicities of the last column. This is clear from figure \ref{ToricE3tophaseIphaseIIC3} where the point ($0,0,0$) correspond to two GLSM fields for phase-I and three GLSM fields for phase-II respectively. We tried removing various points and found that only Fanos $\Ec$ and $\Fa$ are the possible candidates in which Fano $\Cc$ can be embedded. Once again, since we have different number of columns in the two phases of Fano $\Cc$, we will study the embedding and partial resolution of both the phases separately. 
\subsection{Embedding of Fano $\Cc$ inside Fano $\Ec$}
The toric diagram for Fano $\Ec$ is shown in figure \ref{ToricE3tophaseIphaseIIC3}($a$) and can be encoded into the following toric data \cite{Davey:2011mz}: 
\begin{equation}
{\toricG}_{\Ec}=
\left(
\begin{array}{cccccccccc}
 p_1 & p_2 & p_3 & p_4 & p_5 & p_6 & p_7 & p_8 & p_9 & p_{10} \\ \hline
 1 & 1 & 1 & 1 & 1 & 1 & 1 & 1 & 1 & 1 \\
 1 & -1 & 0 & 0 & 0 & 0 & 0 & 0 & 0 & 0 \\
 0 & 0 & 1 & -1 & 1 & 0 & 0 & 0 & 0 & 0 \\
 0 & 0 & 0 & 0 & 1 & 1 & -1 & 0 & 0 & 0
\end{array}
\right)~.
\label{G-E3-first}
\end{equation}
The quiver Chern-Simons theory for Fano $\Ec$ was obtained in \cite{Davey:2011mz} and the charge matrix is given as:
\begin{equation}
Q_{\Ec} = 
\left(
\begin{array}{c}
 Q_F \\ \hline
 Q_D
\end{array}
\right) =
\left(
\begin{array}{cccccccccc}
 1 & 1 & 0 & 0 & 0 & 0 & 0 & -1 & -1 & 0 \\
 0 & 0 & 1 & 1 & 0 & -1 & -1 & 0 & 0 & 0 \\
 0 & 0 & 0 & 1 & 1 & -1 & 0 & 0 & 0 & -1 \\ \hline
 0 & 0 & 0 & 0 & 0 & 1 & 1 & -2 & 0 & 0 \\
 0 & 0 & 0 & 0 & 0 & 0 & 0 & 1 & 0 & -1 \\
 0 & 0 & 0 & 0 & 0 & 0 & 0 & 0 & 1 & -1
\end{array}
\right)~.
\end{equation}
The toric data of both the phases of Fano $\Cc$ are embedded inside the toric data of Fano $\Ec$ (\ref{G-E3-first}) and hence partial resolution of Fano $\Ec$ to get phases of Fano $\Cc$ is possible. We will discuss these in following subsections.

\subsubsection{Partial resolution of Fano $\Ec$ to phase-I of Fano $\Cc$}
The toric data of phase-I of Fano $\Cc$ can be embedded inside the toric data of Fano $\Ec$ (\ref{G-E3-first}), if we remove the set of points $\{p_5,p_8\}$, $\{p_5,p_9\}$ or $\{p_5,p_{10}\}$ (see figure \ref{ToricE3tophaseIphaseIIC3}) as given below:
\begin{eqnarray}
{\cal{G}}^{Phase-I}_{\Cc} = 
\left(
\begin{array}{cccc}
 1 & 0 & 0 & 0 \\
 0 & 1 & 0 & 0 \\
 0 & 0 & 1 & 0 \\
 0 & 0 & 0 & 1
\end{array}
\right).
\left(
\begin{array}{cccccccc}
 p_1 & p_2 & p_3 & p_4 & p_6 & p_7 & p_i & p_{j} \\ \hline
 1 & 1 & 1 & 1 & 1 & 1 & 1 & 1 \\
 1 & -1 & 0 & 0 & 0 & 0 & 0 & 0 \\
 0 & 0 & 1 & -1 & 0 & 0 & 0 & 0 \\
 0 & 0 & 0 & 0 & 1 & -1 & 0 & 0
\end{array}
\right) ~,
\end{eqnarray}
where, $i, j \in (8, 9, 10)$ and are distinct.
The partial resolution method by removing the points mentioned above, gave three reduced charge matrices for Fano $\Cc$ which are all different from the charge matrices $Q^{Phase-I}_{\Cc}$ and $Q^{Phase-II}_{\Cc}$ of the two known quiver Chern-Simons theories for Fano $\Cc$. The one corresponding to the removal of $\{p_5,p_8\}$ and its perfect matching matrix is given as:
\begin{eqnarray}
Q_1 & = & 
\left(
\begin{array}{c}
 Q_F \\ \hline
 Q_D
\end{array}
\right) =
\left(
\begin{array}{cccccccc}
 0 & 0 & 1 & 1 & -1 & -1 & 0 & 0 \\ \hline
 1 & 1 & 0 & 0 & 0 & 0 & -1 & -1 \\
 0 & 0 & 0 & 0 & 1 & 1 & 0 & -2 \\
 0 & 0 & 0 & 0 & 0 & 0 & 1 & -1
\end{array}
\right) ~, \nonumber \\ 
P_1 &=& \left(
\begin{array}{c|cccccccc}
& p_1 & p_2 & p_3 & p_4 & p_5 & p_6 & p_7 & p_8 \\ \hline
X_1 & 1 & 0 & 0 & 0 & 0 & 0 & 0 & 0 \\
X_2 & 0 & 1 & 0 & 0 & 0 & 0 & 0 & 0 \\
X_3 & 0 & 0 & 1 & 0 & 1 & 0 & 0 & 0 \\
X_4 & 0 & 0 & 0 & 1 & 1 & 0 & 0 & 0 \\
X_5 & 0 & 0 & 1 & 0 & 0 & 1 & 0 & 0 \\
X_6 & 0 & 0 & 0 & 1 & 0 & 1 & 0 & 0 \\
X_7 & 0 & 0 & 0 & 0 & 0 & 0 & 1 & 0 \\
X_8 & 0 & 0 & 0 & 0 & 0 & 0 & 0 & 1 \\
\end{array}
\right)~.
\end{eqnarray}
The second reduced charge matrix and its perfect matching matrix obtained by removal of $\{p_5,p_9\}$ is given as:
\begin{eqnarray}
Q_2 &=& 
\left(
\begin{array}{c}
 Q_F \\ \hline
 Q_D
\end{array}
\right) =
\left(
\begin{array}{cccccccc}
 0 & 0 & 1 & 1 & -1 & -1 & 0 & 0 \\ \hline
 1 & 1 & 0 & 0 & 0 & 0 & -1 & -1 \\
 0 & 0 & 0 & 0 & 1 & 1 & -2 & 0 \\
 0 & 0 & 0 & 0 & 0 & 0 & 1 & -1
\end{array}
\right) ~, \nonumber \\
P_2 &=& \left(
\begin{array}{c|cccccccc}
& p_1 & p_2 & p_3 & p_4 & p_5 & p_6 & p_7 & p_8 \\ \hline
X_1 & 1 & 0 & 0 & 0 & 0 & 0 & 0 & 0 \\
X_2 & 0 & 1 & 0 & 0 & 0 & 0 & 0 & 0 \\
X_3 & 0 & 0 & 1 & 0 & 1 & 0 & 0 & 0 \\
X_4 & 0 & 0 & 0 & 1 & 1 & 0 & 0 & 0 \\
X_5 & 0 & 0 & 1 & 0 & 0 & 1 & 0 & 0 \\
X_6 & 0 & 0 & 0 & 1 & 0 & 1 & 0 & 0 \\
X_7 & 0 & 0 & 0 & 0 & 0 & 0 & 1 & 0 \\
X_8 & 0 & 0 & 0 & 0 & 0 & 0 & 0 & 1 \\
\end{array}
\right)~.
\label{Phase-I-C3-INTO-E3-charge-Q2}
\end{eqnarray}
The third and last reduced charge matrix for Fano $\Cc$ was obtained due to removal of points $\{p_5,p_{10}\}$ from the toric diagram for Fano $\Ec$ and is given below:
\begin{eqnarray}
Q_3 &=& 
\left(
\begin{array}{c}
 Q_F \\ \hline
 Q_D
\end{array}
\right) =
\left(
\begin{array}{cccccccc}
 1 & 1 & 0 & 0 & 0 & 0 & -1 & -1 \\
 0 & 0 & 1 & 1 & -1 & -1 & 0 & 0 \\ \hline
 0 & 0 & 0 & 0 & 1 & 1 & -2 & 0 \\
 0 & 0 & 0 & 0 & 0 & 0 & 1 & -1
\end{array}
\right) ~, \nonumber \\
P_3 &=& \left(
\begin{array}{c|cccccccc}
& p_1 & p_2 & p_3 & p_4 & p_5 & p_6 & p_7 & p_8 \\ \hline
X_1 & 0 & 0 & 1 & 0 & 1 & 0 & 0 & 0 \\
X_2 & 0 & 0 & 0 & 1 & 1 & 0 & 0 & 0 \\
X_3 & 0 & 0 & 1 & 0 & 0 & 1 & 0 & 0 \\
X_4 & 0 & 0 & 0 & 1 & 0 & 1 & 0 & 0 \\
X_5 & 1 & 0 & 0 & 0 & 0 & 0 & 1 & 0 \\
X_6 & 0 & 1 & 0 & 0 & 0 & 0 & 1 & 0 \\
X_7 & 1 & 0 & 0 & 0 & 0 & 0 & 0 & 1 \\
X_8 & 0 & 1 & 0 & 0 & 0 & 0 & 0 & 1 \\
\end{array}
\right)~.
\label{Phase-I-C3-INTO-E3-charge-Q3}
\end{eqnarray}
We did not find any quiver diagrams for any of these three $P_1, P_2, P_3$ matrices.

\subsubsection{Partial resolution of Fano $\Ec$ to phase-II of Fano $\Cc$}
Removal of point $p_5$ from the toric data of Fano $\Ec$ (\ref{G-E3-first}) gives the toric data for the phase-II of Fano $\Cc$ (see figure \ref{ToricE3tophaseIphaseIIC3}):
\begin{equation}
{\cal{G}}^{Phase-II}_{\Cc} =
\left(
\begin{array}{cccc}
 1 & 0 & 0 & 0 \\
 0 & 1 & 0 & 0 \\
 0 & 0 & 1 & 0 \\
 0 & 0 & 0 & 1
\end{array}
\right).
\left(
\begin{array}{ccccccccc}
 p_1 & p_2 & p_3 & p_4 & p_6 & p_7 & p_8 & p_9 & p_{10} \\ \hline
 1 & 1 & 1 & 1 & 1 & 1 & 1 & 1 & 1 \\
 1 & -1 & 0 & 0 & 0 & 0 & 0 & 0 & 0 \\
 0 & 0 & 1 & -1 & 0 & 0 & 0 & 0 & 0 \\
 0 & 0 & 0 & 0 & 1 & -1 & 0 & 0 & 0
\end{array}
\right) 
\end{equation}
The partial resolution gives the following reduced charge matrix and perfect matching matrix,
\begin{eqnarray}
Q &=& 
\left(
\begin{array}{c}
 Q_F \\ \hline
 Q_D
\end{array}
\right) =
\left(
\begin{array}{ccccccccc}
 1 & 1 & 0 & 0 & 0 & 0 & -1 & -1 & 0 \\
 0 & 0 & 1 & 1 & -1 & -1 & 0 & 0 & 0 \\ \hline
 0 & 0 & 0 & 0 & 1 & 1 & -2 & 0 & 0 \\
 0 & 0 & 0 & 0 & 0 & 0 & 1 & 0 & -1 \\
 0 & 0 & 0 & 0 & 0 & 0 & 0 & 1 & -1
\end{array}
\right) ~, \nonumber \\
P &=& \left(
\begin{array}{c|ccccccccc}
& p_1 & p_2 & p_3 & p_4 & p_6 & p_7 & p_8 & p_9 \\ \hline
X_1 & 0 & 0 & 1 & 0 & 1 & 0 & 0 & 0 & 0 \\
X_2 & 0 & 0 & 0 & 1 & 1 & 0 & 0 & 0 & 0 \\
X_3 & 0 & 0 & 1 & 0 & 0 & 1 & 0 & 0 & 0 \\
X_4 & 0 & 0 & 0 & 1 & 0 & 1 & 0 & 0 & 0 \\
X_5 & 1 & 0 & 0 & 0 & 0 & 0 & 1 & 0 & 0 \\
X_6 & 0 & 1 & 0 & 0 & 0 & 0 & 1 & 0 & 0 \\
X_7 & 1 & 0 & 0 & 0 & 0 & 0 & 0 & 1 & 0 \\
X_8 & 0 & 1 & 0 & 0 & 0 & 0 & 0 & 1 & 0 \\
X_9 & 0 & 0 & 0 & 0 & 0 & 0 & 0 & 0 & 1 \\
\end{array}
\right)~.
\label{Phase-II-C3-INTO-E3-charge-Q}
\end{eqnarray}
This is again different from any known charge matrices for Fano $\Cc$. But we do not get any quiver diagram from this. 

\subsection{Embedding of Fano $\Cc$ inside Fano $\Fa$}
The toric diagram for Fano $\Fa$ is shown in figure \ref{ToricF1} where the internal point ($0,0,0$) has multiplicity 5. The toric data is given as \cite{Davey:2011mz}:
\begin{figure}[tbp]
	\centering
		\includegraphics[width=0.6\textwidth]{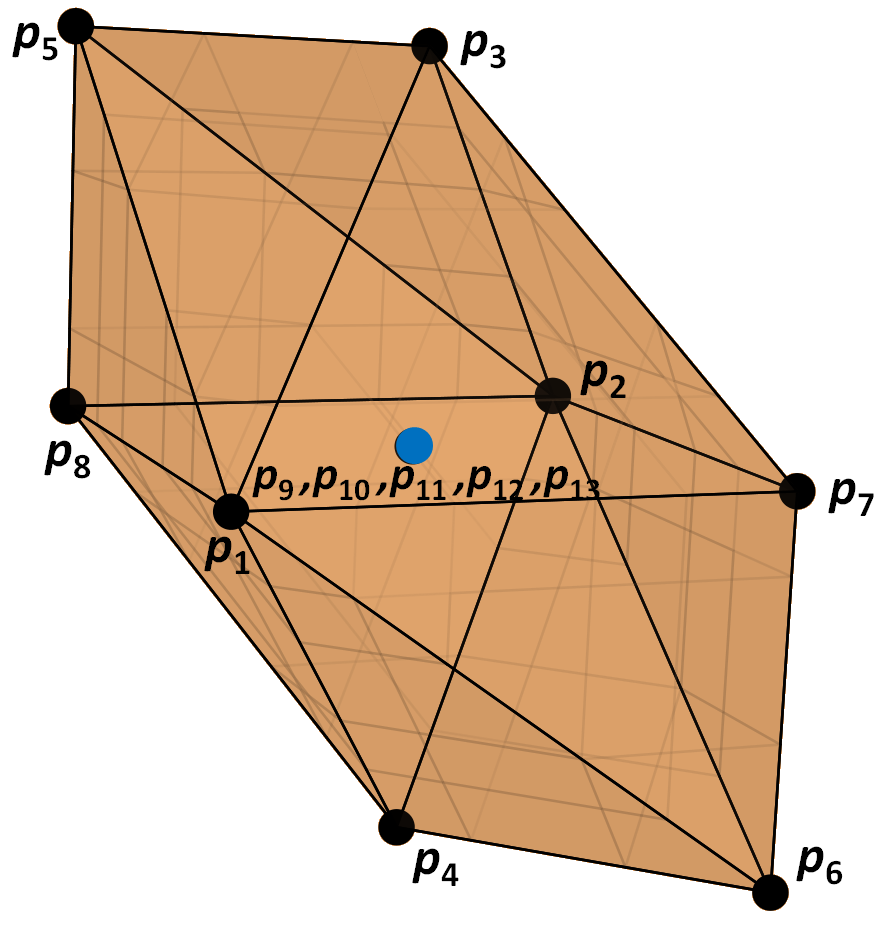}
	\caption{Toric diagram for Fano $\Fa$. The internal point ($0,0,0$) shown in blue correspond to GLSM fields $\{p_9, p_{10}, p_{11}, p_{12}, p_{13}\}$ according to toric data (\ref{G-F1-first}) and thus has multiplicty 5.}
	\label{ToricF1}
\end{figure}
\begin{equation}
{\toricG}_{\Fa} = 
\left(
\begin{array}{ccccccccccccc}
 p_1 & p_2 & p_3 & p_4 & p_5 & p_6 & p_7 & p_8 & p_9 & p_{10} & p_{11} & p_{12} & p_{13} \\ \hline
 1 & 1 & 1 & 1 & 1 & 1 & 1 & 1 & 1 & 1 & 1 & 1 & 1 \\
 1 & -1 & 0 & 0 & 0 & 0 & 0 & 0 & 0 & 0 & 0 & 0 & 0 \\
 0 & 0 & 1 & -1 & 1 & -1 & 0 & 0 & 0 & 0 & 0 & 0 & 0 \\
 0 & 0 & 0 & 0 & 1 & -1 & -1 & 1 & 0 & 0 & 0 & 0 & 0
\end{array}
\right) ~.
\label{G-F1-first}
\end{equation}
The quiver Chern-Simons theory corresponding to Fano $\Fa$ was discussed in \cite{Davey:2011mz} whose charge matrices are given as:
\begin{equation}
Q_{\Fa} = 
\left(
\begin{array}{c}
 Q_F \\ \hline
 Q_D
\end{array}
\right) =
\left(
\begin{array}{ccccccccccccc}
 1 & 1 & 0 & 0 & 0 & 0 & 0 & 0 & -1 & -1 & 0 & 0 & 0 \\
 0 & 0 & 1 & 0 & -1 & 0 & -1 & 0 & 0 & 0 & 0 & 0 & 1 \\
 0 & 0 & 0 & 1 & 0 & -1 & 1 & 0 & 0 & 0 & -1 & 0 & 0 \\
 0 & 0 & 0 & 0 & 1 & 1 & 0 & 0 & 0 & 0 & 0 & -1 & -1 \\
 0 & 0 & 0 & 0 & 0 & 0 & 1 & 1 & 0 & 0 & -1 & 0 & -1 \\ \hline
 0 & 0 & 0 & 0 & 0 & 0 & 0 & 0 & 1 & 0 & -1 & 0 & 0 \\
 0 & 0 & 0 & 0 & 0 & 0 & 0 & 0 & 0 & 1 & 0 & -1 & 0 \\
 0 & 0 & 0 & 0 & 0 & 0 & 0 & 0 & 0 & 0 & 1 & 0 & -1 \\
 0 & 0 & 0 & 0 & 0 & 0 & 0 & 0 & 0 & 0 & 0 & 1 & -1
\end{array}
\right) ~.
\end{equation}
The toric data of both the phases of Fano $\Cc$ can be embedded inside the toric data of Fano $\Fa$. In the following subsections, we will discuss about these embeddings and also list the charge matrices which were obtained as a result of the partial resolution of Fano $\Fa$.

\subsubsection{Partial resolution of Fano $\Fa$ to phase-I of Fano $\Cc$}
The toric data of phase-I of Fano $\Cc$ can be embedded inside the toric data of Fano $\Fa$ (\ref{G-F1-first}) by removing the following set of points: $\{p_5, p_6, p_{9}, p_{10}, p_{11}\}$, $\{p_5, p_6, p_{9}, p_{10}, p_{12}\}$, $\{p_5, p_6, p_{9}, p_{10}, p_{13}\}$, $\{p_5, p_6, p_{9}, p_{11}, p_{12}\}$, $\{p_5, p_6, p_{9}, p_{11}, p_{13}\}$, $\{p_5, p_6, p_{9}, p_{12}, p_{13}\}$ $\{p_5, p_6, p_{10}, p_{11}, p_{12}\}$, $\{p_5, p_6, p_{10}, p_{11}, p_{13}\}$, $\{p_5, p_6, p_{10}, p_{12}, p_{13}\}$, $\{p_5, p_6, p_{11}, p_{12}, p_{13}\}$. Thus, there are 10 different choices of removing points from the toric data of Fano $\Fa$ to get back the toric data of phase-I of Fano $\Cc$. So, partial resolution, in principle, should give 10 possible reduced charge matrices for Fano $\Cc$. However, we found that not all the charge matrices are different. There are only 5 distinct reduced charge matrices $Q_1, Q_2, Q_3, Q_4, Q_5$. All these charge matrices are different from the known charge matrices $Q^{Phase-I}_{\Cc}$ and $Q^{Phase-II}_{\Cc}$ of Fano $\Cc$.

The first reduced charge matrix $Q_1$ was obtained due to removal of $\{p_5, p_6, p_{9}, p_{10}, p_{11}\}$ and is given as:
\begin{equation}
Q_1 = 
Q_D =
\left(
\begin{array}{cccccccc}
 1 & 1 & 0 & 0 & 0 & 0 & -1 & -1 \\
 0 & 0 & 1 & 1 & 0 & 0 & -1 & -1 \\
 0 & 0 & 0 & 0 & 1 & 1 & 0 & -2 \\
 0 & 0 & 0 & 0 & 0 & 0 & 1 & -1
\end{array}
\right) ~.
\end{equation}
The second charge matrix $Q_2$ was obtained from the partial resolution corresponding to the following removal of points: $\{p_5, p_6, p_{9}, p_{11}, p_{12}\}$, $\{p_5, p_6, p_{9}, p_{11}, p_{13}\}$, $\{p_5, p_6, p_{9}, p_{12}, p_{13}\}$, $\{p_5, p_6, p_{10}, p_{11}, p_{12}\}$, $\{p_5, p_6, p_{10}, p_{11}, p_{13}\}$, $\{p_5, p_6, p_{10}, p_{12}, p_{13}\}$. This charge matrix is given as:
\begin{equation}
Q_2 = 
Q_D =
\left(
\begin{array}{cccccccc}
 1 & 1 & 0 & 0 & 0 & 0 & -1 & -1 \\
 0 & 0 & 1 & 1 & 0 & 0 & 0 & -2 \\
 0 & 0 & 0 & 0 & 1 & 1 & 0 & -2 \\
 0 & 0 & 0 & 0 & 0 & 0 & 1 & -1
\end{array}
\right)~.
\end{equation}
The third reduced charge matrix $Q_3$ corresponds to the removal of $\{p_5, p_6, p_{9}, p_{10}, p_{12}\}$ and is given as:
\begin{equation}
Q_3 = 
\left(
\begin{array}{c}
 Q_F \\ \hline
 Q_D
\end{array}
\right) =
\left(
\begin{array}{cccccccc}
 0 & 0 & 0 & 0 & 1 & 1 & -1 & -1 \\ \hline
 1 & 1 & 0 & 0 & 0 & 0 & -1 & -1 \\
 0 & 0 & 1 & 1 & 0 & 0 & -1 & -1 \\
 0 & 0 & 0 & 0 & 0 & 0 & 1 & -1
\end{array}
\right)~.
\label{Phase-I-C3-INTO-F1-charge-Q3}
\end{equation}
The fourth reduced charge matrix $Q_4$ corresponding to removal of $\{p_5, p_6, p_{9}, p_{10}, p_{13}\}$ is given as:
\begin{equation}
Q_4 = 
\left(
\begin{array}{c}
 Q_F \\ \hline
 Q_D
\end{array}
\right) =
\left(
\begin{array}{cccccccc}
 0 & 0 & 1 & 1 & 0 & 0 & -1 & -1 \\ \hline
 1 & 1 & 0 & 0 & 0 & 0 & -1 & -1 \\
 0 & 0 & 0 & 0 & 1 & 1 & -1 & -1 \\
 0 & 0 & 0 & 0 & 0 & 0 & 1 & -1
\end{array}
\right)~.
\label{Phase-I-C3-INTO-F1-charge-Q4}
\end{equation}
The fifth and last reduced charge matrix $Q_5$ was obtained due to removal of $\{p_5, p_6, p_{11}, p_{12}, p_{13}\}$ and is given as:
\begin{equation}
Q_5 = 
\left(
\begin{array}{c}
 Q_F \\ \hline
 Q_D
\end{array}
\right) =
\left(
\begin{array}{cccccccc}
 1 & 1 & 0 & 0 & 0 & 0 & -1 & -1 \\ \hline
 0 & 0 & 1 & 1 & 0 & 0 & -2 & 0 \\
 0 & 0 & 0 & 0 & 1 & 1 & -2 & 0 \\
 0 & 0 & 0 & 0 & 0 & 0 & -1 & 1
\end{array}
\right)~.
\label{Phase-I-C3-INTO-F1-charge-Q5}
\end{equation}
These charge matrices do not give any quiver diagram under inverse algorithm.

\subsubsection{Partial resolution of Fano $\Fa$ to phase-II of Fano $\Cc$}
The toric diagram of phase-II of Fano $\Cc$ can be embedded inside toric diagram of Fano $\Fa$, if remove the following sets of points: $\{p_5, p_6, p_{9}, p_{10}\}$, $\{p_5, p_6, p_{9}, p_{11}\}$, $\{p_5, p_6, p_{9}, p_{12}\}$, $\{p_5, p_6, p_{9}, p_{13}\}$, $\{p_5, p_6, p_{10}, p_{11}\}$, $\{p_5, p_6, p_{10}, p_{12}\}$, $\{p_5, p_6, p_{10}, p_{13}\}$, $\{p_5, p_6, p_{11}, p_{12}\}$, $\{p_5, p_6, p_{11}, p_{13}\}$, $\{p_5, p_6, p_{12}, p_{13}\}$. There are 10 possibilities of removing points, however, partial resolution gives only 8 different reduced charge matrices for Fano $\Cc$. We have listed these 8 charge matrices for Fano $\Cc$ below:
\begin{eqnarray} 
Q_1  & = &
\left(
\begin{array}{c}
 Q_F \\ \hline
 Q_D
\end{array}
\right) = 
\left(
\begin{array}{ccccccccc}
 0 & 0 & 1 & 1 & 0 & 0 & -1 & -1 & 0 \\
 0 & 0 & 0 & 0 & 1 & 1 & -1 & 0 & -1 \\ \hline
 1 & 1 & 0 & 0 & 0 & 0 & -1 & -1 & 0 \\
 0 & 0 & 0 & 0 & 0 & 0 & 1 & 0 & -1 \\
 0 & 0 & 0 & 0 & 0 & 0 & 0 & 1 & -1
\end{array}
\right) ~ \nonumber \\
Q_2  & = & 
\left(
\begin{array}{c}
 Q_F \\ \hline
 Q_D
\end{array}
\right) = 
\left(
\begin{array}{ccccccccc}
 0 & 0 & 1 & 1 & 0 & 0 & 0 & -1 & -1 \\ \hline
 1 & 1 & 0 & 0 & 0 & 0 & -1 & -1 & 0 \\
 0 & 0 & 0 & 0 & 1 & 1 & 0 & -1 & -1 \\
 0 & 0 & 0 & 0 & 0 & 0 & 1 & 0 & -1 \\
 0 & 0 & 0 & 0 & 0 & 0 & 0 & 1 & -1
\end{array}
\right) ~ \nonumber \\
Q_3  & = &  
\left(
\begin{array}{c}
 Q_F \\ \hline
 Q_D
\end{array}
\right) = 
\left(
\begin{array}{ccccccccc}
 0 & 0 & 0 & 0 & 1 & 1 & 0 & -1 & -1 \\ \hline
 1 & 1 & 0 & 0 & 0 & 0 & -1 & 0 & -1 \\
 0 & 0 & 1 & 1 & 0 & 0 & 0 & -1 & -1 \\
 0 & 0 & 0 & 0 & 0 & 0 & 1 & -1 & 0 \\
 0 & 0 & 0 & 0 & 0 & 0 & 0 & 1 & -1
\end{array}
\right)~; \nonumber \\
Q_4   & = &  
\left(
\begin{array}{c}
 Q_F \\ \hline
 Q_D
\end{array}
\right) = 
\left(
\begin{array}{ccccccccc}
 0 & 0 & 1 & 1 & 0 & 0 & 0 & -1 & -1 \\ \hline
 1 & 1 & 0 & 0 & 0 & 0 & -1 & 0 & -1 \\
 0 & 0 & 0 & 0 & 1 & 1 & 0 & -1 & -1 \\
 0 & 0 & 0 & 0 & 0 & 0 & 1 & -1 & 0 \\
 0 & 0 & 0 & 0 & 0 & 0 & 0 & 1 & -1
\end{array}
\right) \nonumber \\
Q_5  & = &  
\left(
\begin{array}{c}
 Q_F \\ \hline
 Q_D
\end{array}
\right) = 
\left(
\begin{array}{ccccccccc}
 1 & 1 & 0 & 0 & 0 & 0 & -1 & -1 & 0 \\ \hline
 0 & 0 & 1 & 1 & 0 & 0 & 0 & 0 & -2 \\
 0 & 0 & 0 & 0 & 1 & 1 & 0 & 0 & -2 \\
 0 & 0 & 0 & 0 & 0 & 0 & 1 & 0 & -1 \\
 0 & 0 & 0 & 0 & 0 & 0 & 0 & 1 & -1
\end{array}
\right) ~; \nonumber \\ 
Q_6  & = & 
\left(
\begin{array}{c}
 Q_F \\ \hline
 Q_D
\end{array}
\right) = 
\left(
\begin{array}{ccccccccc}
 0 & 0 & 0 & 0 & 1 & 1 & 0 & -1 & -1 \\ \hline
 1 & 1 & 0 & 0 & 0 & 0 & -1 & -1 & 0 \\
 0 & 0 & 1 & 1 & 0 & 0 & 0 & -1 & -1 \\
 0 & 0 & 0 & 0 & 0 & 0 & 1 & 0 & -1 \\
 0 & 0 & 0 & 0 & 0 & 0 & 0 & 1 & -1
\end{array}
\right) \nonumber \\
Q_7 & = &
Q_D = 
\left(
\begin{array}{ccccccccc}
 1 & 1 & 0 & 0 & 0 & 0 & -1 & 0 & -1 \\
 0 & 0 & 1 & 1 & 0 & 0 & 0 & -1 & -1 \\
 0 & 0 & 0 & 0 & 1 & 1 & 0 & 0 & -2 \\
 0 & 0 & 0 & 0 & 0 & 0 & 1 & -1 & 0 \\
 0 & 0 & 0 & 0 & 0 & 0 & 0 & 1 & -1
\end{array}
\right) ~; \nonumber \\ 
Q_8 & = &
Q_D = 
\left(
\begin{array}{ccccccccc}
 1 & 1 & 0 & 0 & 0 & 0 & -1 & -1 & 0 \\
 0 & 0 & 1 & 1 & 0 & 0 & 0 & -1 & -1 \\
 0 & 0 & 0 & 0 & 1 & 1 & 0 & 0 & -2 \\
 0 & 0 & 0 & 0 & 0 & 0 & 1 & 0 & -1 \\
 0 & 0 & 0 & 0 & 0 & 0 & 0 & 1 & -1
\end{array}
\right) ~.
\label{Phase-II-C3-INTO-F1-charge-Q1}
\end{eqnarray}
Out of these 8 charge matrices, no diagram is possible for $Q_2, Q_3, Q_4, Q_5, Q_6, Q_7$ and $Q_8$. For $Q_1$, we get 10 different diagrams but they are not quiver diagrams.

\section{Embedding of Fano $\Db$ inside Fano $\Ed$}
\label{sec8}
\begin{figure}[tbp]
	\centering
		\includegraphics[width=0.6\textwidth]{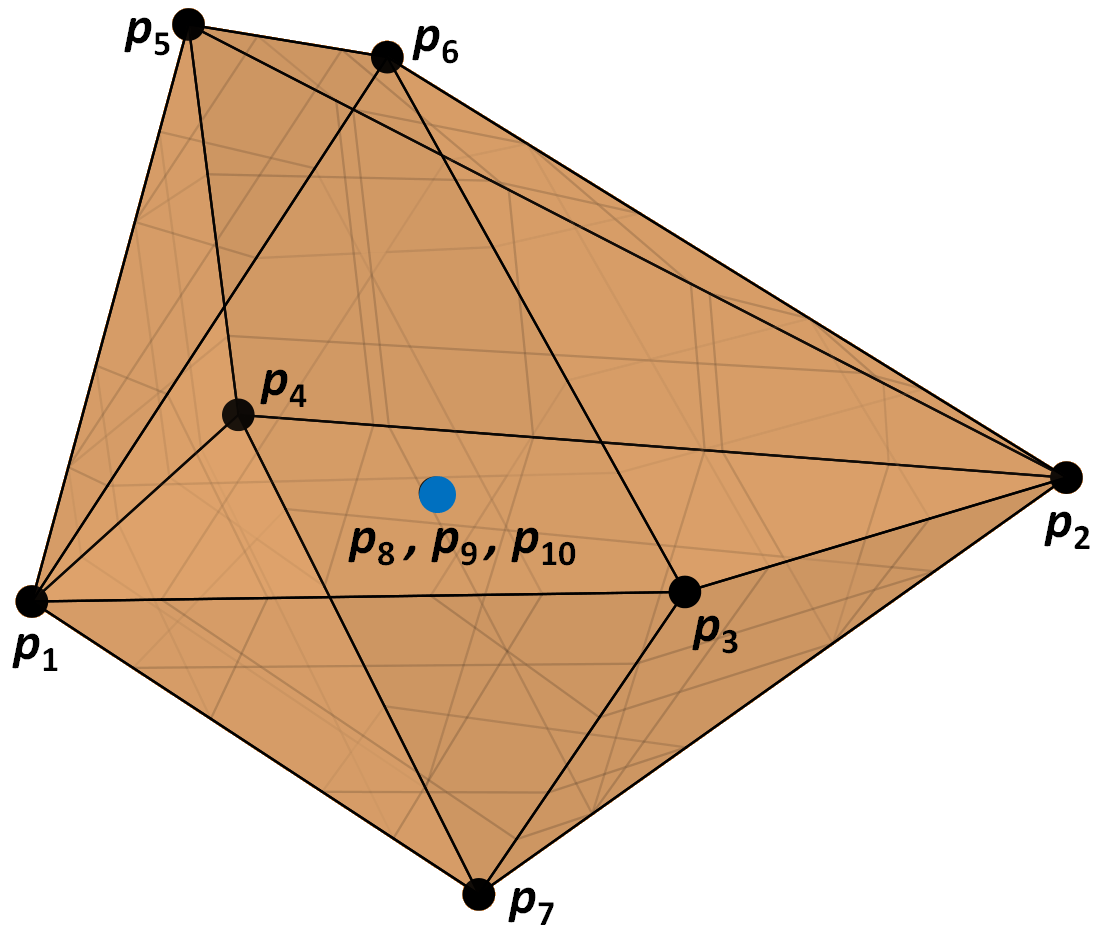}
	\caption{Toric diagram for Fano $\Ed$ with the toric data given in (\ref{G-E4}). The point ($0,0,0$) has multiplicity 3.}
	\label{ToricE4}
\end{figure}

The toric diagram for Fano $\Db$ is shown in figure \ref{ToricD2C4toB4}($a$) and toric data is given in (\ref{G-D2-first}).
The quiver gauge theory for Fano $\Db$ was given in \cite{Davey:2011mz} and the corresponding charge matrix is given as:
\begin{equation}
Q_{\Db} =
\left(
\begin{array}{c}
 Q_F \\ \hline
 Q_D
\end{array}
\right)
=\left(
\begin{array}{cccccccc}
 1 & 1 & 0 & 1 & -1 & 0 & -1 & -1 \\
 0 & 0 & 1 & -1 & 0 & -1 & 1 & 0 \\ \hline
 0 & 0 & 0 & 0 & 1 & 1 & 0 & -2 \\
 0 & 0 & 0 & 0 & 0 & 0 & 1 & -1
\end{array}
\right) ~.
\label{Q-D2}
\end{equation}
The toric data of Fano $\Db$ can only be embedded inside the toric data of Fano $\Ed$. 
The quiver gauge theory corresponding to Fano $\Ed$ is discussed in \cite{Davey:2011mz}. The information about the quiver gauge theory can be encoded in the charge matrix which is given by \cite{Davey:2011mz}:
\begin{equation}
Q_{\Ed} =\left(
\begin{array}{c}
 Q_F \\ \hline
 Q_D
\end{array}
\right)=\left(
\begin{array}{cccccccccc}
 1 & 1 & -1 & 0 & 0 & 0 & 0 & 0 & 0 & -1 \\
 0 & 0 & 0 & 1 & -1 & 0 & -1 & 0 & 1 & 0 \\
 0 & 0 & 0 & 0 & 0 & 1 & 1 & -1 & -1 & 0 \\ \hline
 0 & 0 & 1 & 1 & 0 & 0 & 0 & 0 & 0 & -2 \\
 0 & 0 & 0 & 0 & 0 & 0 & 0 & 1 & -1 & 0 \\
 0 & 0 & 0 & 0 & 0 & 0 & 0 & 0 & 1 & -1
\end{array}
\right) ~.
\label{Q-E4}
\end{equation}
The toric data for Fano $\Ed$ corresponding to this quiver gauge theory is given as \cite{Davey:2011mz}:
\begin{equation}
{\cal{G}}_{\Ed} = \left(
\begin{array}{cccccccccc}
 p_1 & p_2 & p_3 & p_4 & p_5 & p_6 & p_7 & p_8 & p_9 & p_{10} \\ \hline
 1 & 1 & 1 & 1 & 1 & 1 & 1 & 1 & 1 & 1 \\
 1 & -1 & 0 & 0 & 0 & 0 & 0 & 0 & 0 & 0 \\
 0 & 1 & 1 & -1 & -1 & 0 & 0 & 0 & 0 & 0 \\
 0 & 0 & 0 & 0 & 1 & 1 & -1 & 0 & 0 & 0
\end{array}
\right)~,
\label{G-E4}
\end{equation}
and toric diagram is given in figure \ref{ToricE4}.
We find that the toric data of Fano $\Db$ given in (\ref{G-D2-first}) is related by a $GL(4,\BZ)$ transformation to the toric data of Fano $\Ed$ (\ref{G-E4}) if we remove the set of points $\{p_3,p_8\}$ or $\{p_3,p_9\}$ or $\{p_3,p_{10}\}$ as given below:
\begin{eqnarray}
{\cal{G}}_{\Db}  = 
\left(
\begin{array}{cccc}
 1 & 0 & 0 & 0 \\
 0 & 1 & 0 & 0 \\
 0 & 0 & 1 & 0 \\
 0 & 0 & 0 & 1
\end{array}
\right).\left(
\begin{array}{cccccccc}
 p_1 & p_2 & p_4 & p_5 & p_6 & p_7 & p_i & p_j \\ \hline
 1 & 1 & 1 & 1 & 1 & 1 & 1 & 1 \\
 1 & -1 & 0 & 0 & 0 & 0 & 0 & 0 \\
 0 & 1 & -1 & -1 & 0 & 0 & 0 & 0 \\
 0 & 0 & 0 & 1 & 1 & -1 & 0 & 0
\end{array}
\right) ~,   
\end{eqnarray}
where $i,j$ $\in$ $\{8,9,10\}$ such that $i \neq j$.

Thus we have three choices of removing points from the toric diagram and the partial resolution will give three reduced charge matrices $Q_1, Q_2, Q_3$. Removal of points $\{p_3,p_8\}$ and $\{p_3,p_9\}$ will give the following reduced charge matrices for Fano $\Db$ respectively:
\begin{eqnarray}
Q_1 & = & 
\left(
\begin{array}{c}
 Q_F \\ \hline
 Q_D
\end{array}
\right)=\left(
\begin{array}{cccccccc}
 0 & 0 & 1 & -1 & 0 & -1 & 1 & 0 \\ \hline
 1 & 1 & 1 & 0 & 0 & 0 & 0 & -3 \\
 0 & 0 & 0 & 0 & 1 & 1 & -2 & 0 \\
 0 & 0 & 0 & 0 & 0 & 0 & 1 & -1
\end{array}
\right) \nonumber \\ 
Q_2 & = & Q_D = 
\left(
\begin{array}{cccccccc}
 1 & 1 & 1 & 0 & 0 & 0 & 0 & -3 \\
 0 & 0 & 1 & -1 & 0 & -1 & 1 & 0 \\
 0 & 0 & 0 & 0 & 1 & 1 & -2 & 0 \\
 0 & 0 & 0 & 0 & 0 & 0 & 1 & -1
\end{array}
\right) ~.
\end{eqnarray}
Both of these charge matrices $Q_1,Q_2$ for Fano $\Db$ are different from the charge matrix $Q_{\Db}$ (\ref{Q-D2}) of the known quiver gauge theory for Fano $\Db$ and give the following perfect matching matrices,
\begin{equation}
P_1 = \left(
\begin{array}{c|cccccccc}
& p_1 & p_2 & p_3 & p_4 & p_5 & p_6 & p_7 & p_8 \\ \hline 
X_1 & 1 & 0 & 0 & 0 & 0 & 0 & 0 & 0 \\
X_2 & 0 & 1 & 0 & 0 & 0 & 0 & 0 & 0 \\
X_3 & 0 & 0 & 1 & 1 & 0 & 0 & 0 & 0 \\
X_4 & 0 & 0 & 0 & 0 & 1 & 0 & 0 & 0 \\
X_5 & 0 & 0 & 1 & 0 & 0 & 1 & 0 & 0 \\
X_6 & 0 & 0 & 0 & 1 & 0 & 0 & 1 & 0 \\
X_7 & 0 & 0 & 0 & 0 & 0 & 1 & 1 & 0 \\
 X_8 &0 & 0 & 0 & 0 & 0 & 0 & 0 & 1 \\
\end{array}
\right) ~; P_2 = I_8 ~,
\end{equation}
where $I_8$ is identity matrix of order 8. However we checked that none of them give any quiver diagram.
  
The third choice of removing points $\{p_3,p_{10}\}$ gives the reduced charge matrix $Q_3$ and perfect matching matrix $P_3$ given as:
\begin{eqnarray}
Q_3 &=&\left(
\begin{array}{c}
 Q_F \\ \hline
 Q_D
\end{array}
\right)=\left(
\begin{array}{cccccccc}
 0 & 0 & 1 & -1 & 0 & -1 & 0 & 1 \\
 0 & 0 & 0 & 0 & 1 & 1 & -1 & -1 \\ \hline
 1 & 1 & 1 & 0 & 0 & 0 & 0 & -3 \\
 0 & 0 & 0 & 0 & 0 & 0 & 1 & -1
\end{array}
\right) ~, \nonumber \\
P_3 &=& \left(
\begin{array}{c|cccccccc}
& p_1 & p_2 & p_3 & p_4 & p_5 & p_6 & p_7 & p_8 \\ \hline 
X_1 & 1 & 0 & 0 & 0 & 0 & 0 & 0 & 0 \\
X_2 & 0 & 1 & 0 & 0 & 0 & 0 & 0 & 0 \\
X_3 & 0 & 0 & 1 & 1 & 0 & 0 & 0 & 0 \\
X_4 & 0 & 0 & 0 & 0 & 1 & 0 & 1 & 0 \\
X_5 & 0 & 0 & 1 & 0 & 0 & 1 & 1 & 0 \\
X_6 & 0 & 0 & 0 & 1 & 1 & 0 & 0 & 1 \\
X_7 & 0 & 0 & 0 & 0 & 0 & 1 & 0 & 1 \\
\end{array}
\right)~.
\label{Q-reduced-D2}
\end{eqnarray}
This reduced charge matrix is again different from the charge matrix $Q_{\Db}$ (\ref{Q-D2}) of Fano $\Db$. The $P_3$ matrix gives 12 diagrams but these diagrams are not quiver diagrams.

\section{Embedding of Fano $\Ea$ inside Fano $\Fb$}
\label{sec9}
The toric diagram for Fano $\Ea$ is given in figure \ref{ToricF2toE1}($b$) with toric data given as \cite{Davey:2011mz}:
\begin{figure}[tbp]
	\centering
		\includegraphics[width=1.05\textwidth]{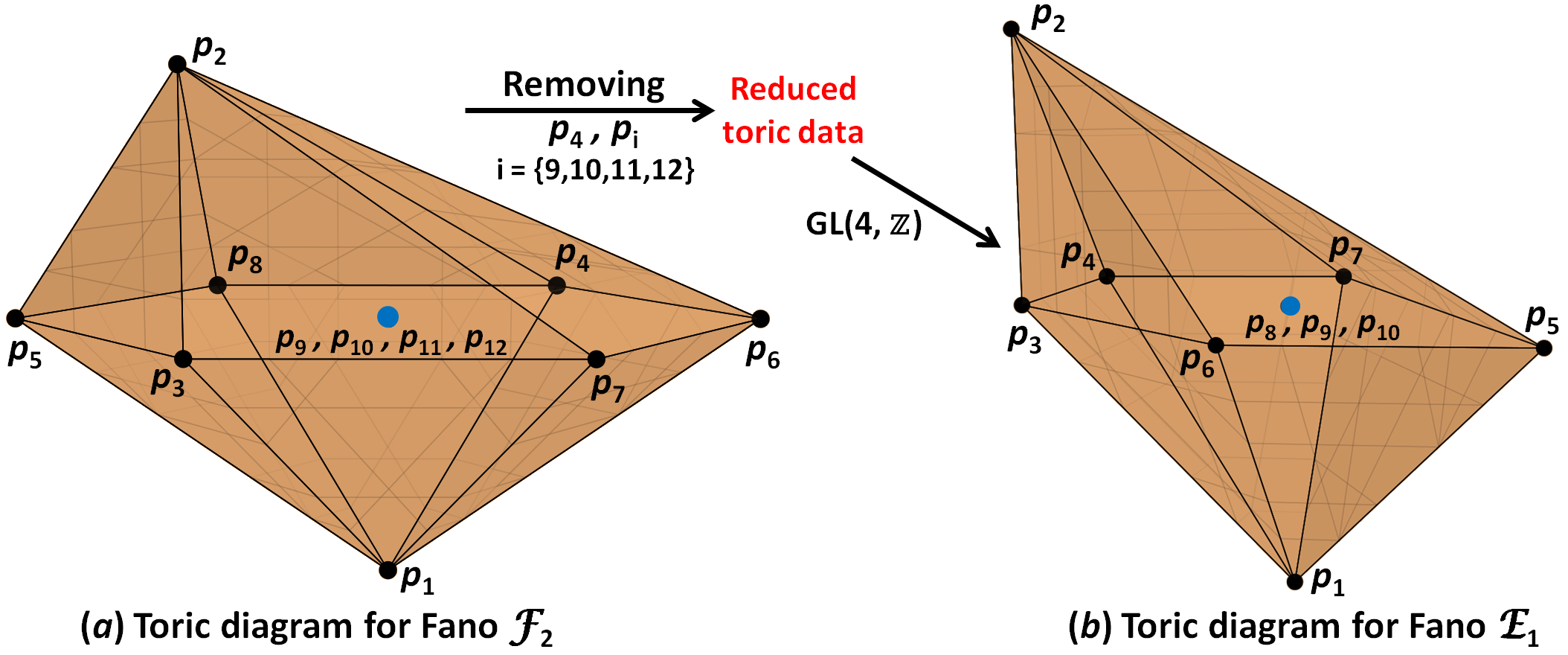}
	\caption{Figure ($a$) shows the toric diagram for Fano $\Fb$ with toric data given in (\ref{G-F2}). Removing $p_4$ and any one of the $\{p_9,p_{10},p_{11},p_{12}\}$ will give a reduced toric data which is equivalent to toric diagram for Fano $\Ea$ with toric data given in (\ref{G-E1}).}
	\label{ToricF2toE1}
\end{figure}
\begin{equation}
{\cal{G}}_{\Ea} = \left(
\begin{array}{cccccccccc}
p_1 & p_2 & p_3 & p_4 & p_5 & p_6 & p_7 & p_8 & p_9 & p_{10} \\ \hline
 1 & 1 & 1 & 1 & 1 & 1 & 1 & 1 & 1 & 1 \\
 1 & -1 & 0 & 0 & 0 & 0 & 0 & 0 & 0 & 0 \\
 0 & 1 & 1 & 1 & -1 & 0 & 0 & 0 & 0 & 0 \\
 0 & 0 & 0 & 1 & -1 & -1 & 1 & 0 & 0 & 0
\end{array}
\right)~.
\label{G-E1}
\end{equation}
The quiver gauge theory for Fano $\Ea$ was given in \cite{Davey:2011mz} and the corresponding charge matrix is given as:
\begin{equation}
Q_{\Ea} =
\left(
\begin{array}{c}
 Q_F \\ \hline
 Q_D
\end{array}
\right)=
\left(
\begin{array}{cccccccccc}
 1 & 1 & 0 & 0 & 1 & -1 & 0 & 0 & -1 & -1 \\
 0 & 0 & 1 & -1 & 0 & 0 & 1 & -1 & 0 & 0 \\
 0 & 0 & 0 & 1 & 1 & -1 & -1 & 0 & 0 & 0 \\ \hline
 0 & 0 & 0 & 0 & 0 & 1 & 1 & 0 & 0 & -2 \\
 0 & 0 & 0 & 0 & 0 & 0 & 0 & 1 & 0 & -1 \\
 0 & 0 & 0 & 0 & 0 & 0 & 0 & 0 & 1 & -1
\end{array}
\right) ~.
\label{Q-E1}
\end{equation}
The embedding of Fano $\Ea$ inside Fano $\Fb$ was shown in \cite{Phukon:2011hp}. We find that the toric diagram of Fano $\Ea$ can only be embedded inside the toric diagram of Fano $\Fb$. 
The quiver gauge theory corresponding to Fano $\Fb$ is discussed in \cite{Davey:2011mz}. The information about the quiver gauge theory can be encoded in the charge matrix which is given by \cite{Davey:2011mz}:
\begin{equation}
Q_{\Fb} =
\left(
\begin{array}{c}
 Q_F \\ \hline
 Q_D
\end{array}
\right) = 
\left(
\begin{array}{cccccccccccc}
 1 & 1 & 0 & 0 & 0 & 1 & -1 & 0 & 0 & 0 & -1 & -1 \\
 0 & 0 & 1 & 0 & 0 & 1 & -1 & 0 & 0 & -1 & 0 & 0 \\
 0 & 0 & 0 & 1 & 0 & -1 & 1 & 0 & -1 & 0 & 0 & 0 \\
 0 & 0 & 0 & 0 & 1 & 1 & -1 & -1 & 0 & 0 & 0 & 0 \\ \hline
 0 & 0 & 0 & 0 & 0 & 0 & 1 & 1 & 0 & 0 & -2 & 0 \\
 0 & 0 & 0 & 0 & 0 & 0 & 0 & 0 & 1 & 0 & 0 & -1 \\
 0 & 0 & 0 & 0 & 0 & 0 & 0 & 0 & 0 & 1 & 0 & -1 \\
 0 & 0 & 0 & 0 & 0 & 0 & 0 & 0 & 0 & 0 & 1 & -1
\end{array}
\right) ~.
\label{Q-F2}
\end{equation}
The toric data for Fano $\Fb$ corresponding to this quiver gauge theory is given as \cite{Davey:2011mz}:
\begin{equation}
{\cal{G}}_{\Fb} = \left(
\begin{array}{cccccccccccc}
 p_1 & p_2 & p_3 & p_4 & p_5 & p_6 & p_7 & p_8 & p_9 & p_{10} & p_{11} & p_{12} \\ \hline
 1 & 1 & 1 & 1 & 1 & 1 & 1 & 1 & 1 & 1 & 1 & 1 \\
 1 & -1 & 0 & 0 & 0 & 0 & 0 & 0 & 0 & 0 & 0 & 0 \\
 0 & 1 & 1 & -1 & 1 & -1 & 0 & 0 & 0 & 0 & 0 & 0 \\
 0 & 0 & 0 & 0 & 1 & -1 & -1 & 1 & 0 & 0 & 0 & 0
\end{array}
\right)~,
\label{G-F2}
\end{equation}
and toric diagram is shown in figure \ref{ToricF2toE1}($a$). The toric data of Fano $\Ea$ given in (\ref{G-E1}) is related by a $GL(4,\BZ)$ transformation to the toric data of Fano $\Fb$ (\ref{G-F2}) after removing the following sets of points: $\{p_4,p_9\}$, $\{p_4,p_{10}\}$, $\{p_4,p_{11}\}$, $\{p_4,p_{12}\}$ as given below:
\begin{eqnarray}
{\cal{G}}_{\Ea}  = 
\left(
\begin{array}{cccc}
 1 & 0 & 0 & 0 \\
 0 & 1 & 0 & 0 \\
 0 & 0 & 1 & 0 \\
 0 & 0 & 0 & 1
\end{array}
\right).\left(
\begin{array}{cccccccccc}
 p_1 & p_2 & p_3 & p_5 & p_6 & p_7 & p_8 & p_i & p_j & p_k \\ \hline
 1 & 1 & 1 & 1 & 1 & 1 & 1 & 1 & 1 & 1 \\
 1 & -1 & 0 & 0 & 0 & 0 & 0 & 0 & 0 & 0 \\
 0 & 1 & 1 & 1 & -1 & 0 & 0 & 0 & 0 & 0 \\
 0 & 0 & 0 & 1 & -1 & -1 & 1 & 0 & 0 & 0
\end{array}
\right) ~,   
\end{eqnarray}
where $i,j,k$ $\in$ $\{9,10,11,12\}$ and take distinct values.

Thus we have four choices of removing points from the toric diagram and the partial resolution, in principle, should give four reduced charge matrices. However we found that removal of points $\{p_4,p_{11}\}$ and $\{p_4,p_{12}\}$ give the same reduced charge matrix. Thus, resolving the Fano $\Fb$ theory, we get three different charge matrices $Q_1, Q_2, Q_3$ for Fano $\Ea$ which are given as:
\begin{eqnarray}
Q_1 & = & 
\left(
\begin{array}{c}
 Q_F \\ \hline
 Q_D
\end{array}
\right)=\left(
\begin{array}{cccccccccc}
 1 & 1 & 0 & 0 & 1 & -1 & 0 & 0 & -1 & -1 \\
 0 & 0 & 1 & 0 & 1 & -1 & 0 & -1 & 0 & 0 \\
 0 & 0 & 0 & 1 & 1 & -1 & -1 & 0 & 0 & 0 \\ \hline
 0 & 0 & 0 & 0 & 0 & 1 & 1 & 0 & -2 & 0 \\
 0 & 0 & 0 & 0 & 0 & 0 & 0 & 1 & 0 & -1 \\
 0 & 0 & 0 & 0 & 0 & 0 & 0 & 0 & 1 & -1
\end{array}
\right)~, \nonumber \\ 
Q_2 & = &
\left(
\begin{array}{c}
 Q_F \\ \hline
 Q_D
\end{array}
\right) = 
\left(
\begin{array}{cccccccccc}
 1 & 1 & 0 & 0 & 1 & -1 & 0 & 0 & -1 & -1 \\
 0 & 0 & 0 & 1 & 1 & -1 & -1 & 0 & 0 & 0 \\ \hline
 0 & 0 & 1 & 0 & 1 & -1 & 0 & 0 & 0 & -1 \\
 0 & 0 & 0 & 0 & 0 & 1 & 1 & 0 & -2 & 0 \\
 0 & 0 & 0 & 0 & 0 & 0 & 0 & 1 & 0 & -1 \\
 0 & 0 & 0 & 0 & 0 & 0 & 0 & 0 & 1 & -1
\end{array}
\right)~, \nonumber \\
Q_3 & = &
\left(
\begin{array}{c}
 Q_F \\ \hline
 Q_D
\end{array}
\right) =
\left(
\begin{array}{cccccccccc}
 0 & 0 & 1 & 0 & 1 & -1 & 0 & 0 & -1 & 0 \\
 0 & 0 & 0 & 1 & 1 & -1 & -1 & 0 & 0 & 0 \\ \hline
 1 & 1 & 0 & 0 & 1 & -1 & 0 & 0 & 0 & -2 \\
 0 & 0 & 0 & 0 & 0 & 1 & 1 & 0 & 0 & -2 \\
 0 & 0 & 0 & 0 & 0 & 0 & 0 & 1 & 0 & -1 \\
 0 & 0 & 0 & 0 & 0 & 0 & 0 & 0 & 1 & -1
\end{array}
\right)~.
\end{eqnarray}
We can see that all these charge matrices $Q_1,Q_2,Q_3$ for Fano $\Ea$ are different from the charge matrix $Q_{\Ea}$ (\ref{Q-E1}) of the known quiver gauge theory for Fano $\Ea$. The perfect matching matrix for $Q_1$ is given as,
\begin{equation}
P_1 = \left(
\begin{array}{c|cccccccccc}
& p_1 & p_2 & p_3 & p_4 & p_5 & p_6 & p_7 & p_8 & p_9 & p_{10} \\ \hline
X_1 & 1 & 0 & 1 & 1 & 0 & 1 & 0 & 0 & 0 & 0 \\
X_2 & 0 & 1 & 1 & 1 & 0 & 1 & 0 & 0 & 0 & 0 \\
X_3 & 0 & 0 & 0 & 0 & 1 & 1 & 0 & 0 & 0 & 0 \\
X_4 & 0 & 0 & 0 & 1 & 0 & 0 & 1 & 0 & 0 & 0 \\
X_5 & 0 & 0 & 1 & 0 & 0 & 0 & 0 & 1 & 0 & 0 \\
X_6 & 1 & 0 & 0 & 0 & 0 & 0 & 0 & 0 & 1 & 0 \\
X_7 & 0 & 1 & 0 & 0 & 0 & 0 & 0 & 0 & 1 & 0 \\
X_8 & 0 & 0 & 0 & 0 & 1 & 0 & 1 & 1 & 1 & 0 \\
X_9 & 1 & 0 & 0 & 0 & 0 & 0 & 0 & 0 & 0 & 1 \\
X_{10} & 0 & 1 & 0 & 0 & 0 & 0 & 0 & 0 & 0 & 1 \\
X_{11} & 0 & 0 & 0 & 0 & 1 & 0 & 1 & 1 & 0 & 1 \\
\end{array}
\right)~.
\end{equation}
However we checked that this matrix does not give any possible quiver diagram. The matching matrix corresponding to $Q_2$ is,
\begin{equation}
P_2 = \left(
\begin{array}{c|cccccccccc}
& p_1 & p_2 & p_3 & p_4 & p_5 & p_6 & p_7 & p_8 & p_9 & p_{10} \\ \hline
X_1 & 0 & 0 & 1 & 0 & 0 & 0 & 0 & 0 & 0 & 0 \\
X_2 & 1 & 0 & 0 & 1 & 0 & 1 & 0 & 0 & 0 & 0 \\
X_3 & 0 & 1 & 0 & 1 & 0 & 1 & 0 & 0 & 0 & 0 \\
X_4 & 0 & 0 & 0 & 0 & 1 & 1 & 0 & 0 & 0 & 0 \\
X_5 & 0 & 0 & 0 & 1 & 0 & 0 & 1 & 0 & 0 & 0 \\
X_6 & 0 & 0 & 0 & 0 & 0 & 0 & 0 & 1 & 0 & 0 \\
X_7 & 1 & 0 & 0 & 0 & 0 & 0 & 0 & 0 & 1 & 0 \\
X_8 & 0 & 1 & 0 & 0 & 0 & 0 & 0 & 0 & 1 & 0 \\
X_9 & 0 & 0 & 0 & 0 & 1 & 0 & 1 & 0 & 1 & 0 \\
X_{10} & 1 & 0 & 0 & 0 & 0 & 0 & 0 & 0 & 0 & 1 \\
X_{11} & 0 & 1 & 0 & 0 & 0 & 0 & 0 & 0 & 0 & 1 \\
X_{12} & 0 & 0 & 0 & 0 & 1 & 0 & 1 & 0 & 0 & 1 \\
\end{array}
\right)~,
\end{equation}
which can not be encoded by any quiver diagram. For reduced charge matrix $Q_3$, the perfect matching matrix comes out to be,
\begin{equation}
P_3 = \left(
\begin{array}{c|cccccccccc}
& p_1 & p_2 & p_3 & p_4 & p_5 & p_6 & p_7 & p_8 & p_9 & p_{10} \\ \hline
X_1 & 1 & 0 & 0 & 0 & 0 & 0 & 0 & 0 & 0 & 0 \\
X_2 & 0 & 1 & 0 & 0 & 0 & 0 & 0 & 0 & 0 & 0 \\
X_3 & 0 & 0 & 1 & 1 & 0 & 1 & 0 & 0 & 0 & 0 \\
X_4 & 0 & 0 & 0 & 0 & 1 & 1 & 0 & 0 & 0 & 0 \\
X_5 & 0 & 0 & 0 & 1 & 0 & 0 & 1 & 0 & 0 & 0 \\
X_6 & 0 & 0 & 0 & 0 & 0 & 0 & 0 & 1 & 0 & 0 \\
X_7 & 0 & 0 & 1 & 0 & 0 & 0 & 0 & 0 & 1 & 0 \\
X_8 & 0 & 0 & 0 & 0 & 1 & 0 & 1 & 0 & 1 & 0 \\
X_9 & 0 & 0 & 0 & 0 & 0 & 0 & 0 & 0 & 0 & 1 \\
\end{array}
\right)~,
\end{equation}
but this matrix too does not give any quiver diagram.

\section{Conclusion}
\label{sec10}
In this work, we have studied the embeddings of toric Calabi-Yau fourfolds which are complex cones over the Fano threefolds. There are 18 Fano threefolds listed in the literature. Our main focus was to find the embeddings of these Fanos inside other Fano threefolds. By removing various possible points from the toric diagrams of these Fano threefolds, we were able to find the following embeddings: Fano $\BP^3$ is embedded inside Fano $\Bc$, Fano $\Bb$ and Fano $\Ce$; Fano $\Ba$ inside Fano $\Cb$; Fano $\Bd$ inside Fano $\Db$ and Fano $\Cd$; Fano $\Ca$ inside Fano $\Eb$; Fano $\Cc$ is embedded inside both Fano $\Ec$ and Fano $\Fa$; Fano $\Db$ inside Fano $\Ed$; Fano $\Ea$ inside Fano $\Fb$. Some of these embeddings like, Fano $\BP^3$ inside Fano $\Bc$ and Fano $\Bb$; Fano $\Bd$ inside Fano $\Db$ and Fano $\Cd$; Fano $\Cc$ inside Fano $\Ec$; Fano $\Ea$ inside Fano $\Fb$, were already known. 

In all these cases, we studied the partial resolution, in order to find new quiver Chern-Simons theories, which may provide new examples of toric dualities. We found some diagrams for Fano $\BP^3$, Fano $\Ba$, Fano $\Bd$, Fano $\Cc$ and Fano $\Db$, using the partial resolution and inverse algorithm approach. But these diagrams do not have equal number of incoming and outgoing arrows at every node and hence are not the quiver diagrams.

By doing the partial resolution of Fano $\Db$, we found a quiver Chern-Simons theory corresponding to Fano $\Bd$ shown in the quiver diagram \ref{D2-INTO-B4}. This theory matches with the already known quiver gauge theory \cite{Davey:2011mz} for Fano $\Bd$. In \cite{Davey:2011mz}, the quiver gauge theory given by quiver \ref{D2-INTO-B4}, was obtained from the brane tiling, and then the forward algorithm was used to verify that this theory corresponds to Fano $\Bd$. In this work, we have obtained the same theory for Fano $\Bd$ via the partial resolution of Fano $\Db$.

We would like to mention that in this work, we have only studied the embeddings of Fano threefolds inside other Fanos. It would be quite interesting to explore the embeddings of these 18 Fanos inside the toric Calabi-Yau fourfolds which are not complex cones over Fano threefolds, and whose quiver Chern-Simons theories are known. In such a case, partial resolution can be used to find new quiver Chern-Simons theories for the Fano threefolds. This may lead to the discovery of new toric dual theories. 

\vskip.5cm
\noindent
{\bf Acknowledgments}: SD would like to thank Tapobrata Sarkar and  
P. Ramadevi for their valuable suggestions regarding this work.

\bibliographystyle{JHEP}
\bibliography{Resolving-Fanos-Biblio}



\end{document}